\documentclass[a5paper, 10pt, fleqn]{scrartcl}
\usepackage[british]{babel}
\usepackage[T1]{fontenc}
\usepackage{graphicx,amsmath,amsfonts,amssymb,microtype,mathdefs,array,mathrsfs,amsthm,enum,relsize}
\usepackage[section]{thmdefs}

\setkomafont{pagehead}{\small\itshape}
\setkomafont{sectioning}{\fontshape{ssc}\selectfont}
\setkomafont{descriptionlabel}{\itshape}

\typearea{calc}

\makeatletter
\newcommand\m@thsm@ller[2]{\mbox{\relscale{0.91}$\m@th#1#2$}}
\let\smaller\undefined
\DeclareRobustCommand\smaller[1]{\relax\ifmmode{\mathpalette\m@thsm@ller{#1}}\else{\relscale{0.91}#1}\fi}
\makeatother

\DeclareRobustCommand*{\dom}{\qopname\relax o{dom}}
\DeclareRobustCommand*{\rng}{\qopname\relax o{rng}}
\newcommand*{\id}{\mathrm{id}}
\newcommand*{\pPos}{\mathsf{pPos}}
\newcommand*{\Pos}{\mathsf{Pos}}
\newcommand*{\Alg}{\mathsf{Alg}}
\newcommand*{\pAlg}{\mathsf{pAlg}}
\newcommand*{\BAlg}{\mathsf{BAlg}}
\newcommand*{\CAlg}{\mathsf{CAlg}}
\newcommand*{\SGrp}{\mathsf{SGrp}}

\newcommand*{\ar}{\mathrm{ar}}
\newcommand*{\sh}{\mathrm{sh}}
\newcommand*{\hole}{\mathrm{hole}}
\newcommand*{\Hole}{\mathrm{Hole}}

\newcommand*{\Flat}{\mathrm{flat}}
\newcommand*{\sing}{\mathrm{sing}}
\newcommand*{\dist}{\mathrm{dist}}
\newcommand*{\union}{\mathrm{union}}

\newcommand*{\pt}{\mathrm{pt}}

\newcommand*{\reg}{\mathrm{reg}}

\newcommand*{\cy}{\mathrm{cy}}

\newcommand*{\Branch}{\mathrm{Branch}}

\newcommand*{\TS}{\mathrm{TS}}
\newcommand*{\SG}{\mathrm{SG}}
\newcommand*{\TA}{\mathrm{TA}}

\newcommand*{\op}{\mathrm{op}}

\newcommand*{\cl}{\mathrm{cl}}
\newcommand*{\un}{\mathrm{un}}

\newcommand*{\Tr}{\mathrm{Tr}}

\newcommand*{\MSO}{\smaller{\mathrm{MSO}}}

\newcommand*{\pf}{\mathrm{pf}}

\DeclareRobustCommand*{\Tr}{\qopname\relax o{Tr}}

\newcommand*{\emptyseq}{\langle\rangle}

\newcommand*{\?}{\kern .08em}
\DeclareRobustCommand*{\Belowseg}{\mathord\Downarrow}
\DeclareRobustCommand*{\Aboveseg}{\mathord\Uparrow}

\newcommand\upqed{\vskip-\baselineskip\vskip-\belowdisplayskip}
\newcommand\markenddef{\hfill$\lrcorner$}

\begin{document}
\title{Branch-Continuous Tree Algebras}
\author{Achim Blumensath}
\maketitle

\tableofcontents

\section{Introduction}   

Algebraic language theory uses tools from algebra to study regular languages.
It has been particularly successful in deriving decidable
characterisations for various fragments of monadic second-order logic.
For instance, a Theorem of Sch\"utzenberger~\cite{Schutzenberger65} states
that a regular language is first-order definable if, and only if,
its syntactic monoid is aperiodic.
The latter condition is decidable as we can compute this syntactic monoid
from an automaton for the language and then check it for aperiodicity.

In recent years there has been an effort to extend this algebraic approach
to languages of infinite trees.
Preliminary results were provided by the group of Boja\'nczyk
\cite{BojanczykId09,BojanczykIdSk13} with one article considering languages of
\emph{regular} trees only, and one considering languages of \emph{thin} trees.
The first complete framework that could deal with arbitrary infinite trees
was provided by Blumensath~\cite{Blumensath11c,Blumensath13a}.
Unfortunately, it turned out to be too complicated and technical for applications.

An interesting new approach has recently been suggested by Blumensath, Boja\'nczyk,
and Klin~\cite{BlumensathBojanczykKlinXX}.
They introduced the class of \emph{regular tree algebras} and showed that
this class characterises the class of all regular languages of infinite trees
in the sense that a tree language is regular if, and only if, it is recognised
by such an algebra. Furthermore, they proved the existence of syntactic algebras
and showed that these algebras are regular. This is all that is required of a
framework if one wants to use it for obtaining decision procedures.
From a theoretical perspective though, the notion of a regular tree algebra
has a serious drawback\?: the definition is circular in the sense that
it is based on the notion of a regular language.
Hence, the framework cannot be used as a replacement for other formalisms
such as automata or logic, as at least one them is required during the development
of the theory of regular tree algebras.
It would be very desirable to have an alternative, purely algebraic definition
of the notion of a regular tree algebra.
Unfortunately, none has been proposed so far.

In this article we introduce a new class of tree algebras,
the so-called \emph{branch-continuous tree algebras,} that also characterises
the class of regular tree languages. The definition is purely algebraic and
therefore does not suffer from the above problems.
In particular, branch-continuous algebras
seem to be a suitable replacement for automata and logic.
On the downside, the new class does not have the nice closure properties
enjoyed by the regular tree algebras. In particular, syntactic algebras are not
necessarily branch-continuous.
Consequently, regular tree algebras seem to be more suited for practical applications,
while the branch-continuous one we consider in the present article appear
to be more useful for developing an algebraic theory of regular languages,
in particular, as far as the study of algebraic and combinatorical properties
of such languages is concerned.

The outline of the article is as follows.
We start in Section~\ref{Sect: tree algebras} with some basic definitions
including that of a tree algebra.
And we begin to develop the algebraic theory of such algebras by introducing some
basic notations such as completeness and continuity.
In Section~\ref{Sect: power-set} we study two completion operations for tree algebras
based on the power-set construction.
This will pay off later on by allowing several proofs to be quite streamlined and concise.

The heart of this article is Section~\ref{Sect: branch-continuous algebras}
where we introduce the central notion of a \emph{branch-continuous}
tree algebra and we prove that they characterise the class of regular languages.
Finally, the last section studies finite representations of branch-continuous
tree algebras by introducing an analogue to Wilke algebras in our setting.

\smallskip
\noindent
\textbf{Acknowledgements.}
Many of the central ideas of this article were developed in discussions with
Thomas Colcombet during a stay in Paris some years ago.
Without him the theory presented here would be much more convoluted.

\section{Tree algebras}   
\label{Sect: tree algebras}

One problem with the old framework of Blumensath~\cite{Blumensath11c,Blumensath13a}
was the complicated notation for the algebras used.
In the meantime a very clean alternative has been proposed which we will adopt
for this article. This alternative is based on the category-theoretical notion
of a monad and an Eilenberg--Moore algebra.
As an example, let us show how define semigroups in this setting.

Given a semigroup $\frakS = \langle S, {}\cdot{}\rangle$, we can extend
the binary product ${}\cdot{} : S^2 \to S$ to a product operation
$\pi : S^+ \to S$ that takes an arbitrary finite sequence of semigroup elements
as argument.
Hence, we can formalise semigroups as structures of the form $\langle S,\pi\rangle$
where $\pi : S^+ \to S$ is an associative operation from the \emph{free semigroup}
generated by~$S$ to~$S$. Associativity in this context means that,
given finite sequences $w_0,\dots,w_{n-1} \in S^+$, we have
\begin{align*}
  \pi(\pi(w_0),\dots,\pi(w_{n-1})) = \pi(w_0\dots w_{n-1})\,.
\end{align*}
Besides associativity we need one additional axiom,
when using a variable-arity product~$\pi$\?:
we have to require that the product of a single element returns that element.
\begin{align*}
  \pi(\langle a\rangle) = a\,, \quad\text{for } a \in S\,.
\end{align*}
Then it follows that every pair $\langle S,\pi\rangle$ satisfying these two axioms
corresponds to a semigroup $\langle S,{}\cdot{}\rangle$ and vice versa.

This point of view can easily be generalised to other kinds of associative algebras.
The only thing we need is the notion of a free algebra generated by some set~$X$.
So, suppose we have a functor~$\bbT$ mapping a set~$X$
to the free algebra~$\bbT X$ generated by~$X$.
Then we can define an algebra as a pair $\langle A,\pi\rangle$ consisting of a
set~$A$ and a product function $\pi : \bbT A \to A$.
To express our two axioms for such an algebra, we also need functions
$\Flat : \bbT\bbT A \to \bbT A$ and $\sing : A \to \bbT A$
that generalise the concatenation and singleton operations
\begin{align*}
  (S^+)^+ \to S^+ &: \langle w_0,\dots,w_{n-1}\rangle \mapsto w_0\dots w_{n-1} \\
  S \to S^+ &: a \mapsto \langle a\rangle
\end{align*}
in the semigroup case.
Then we can write the associativity axiom as

\noindent
\begin{minipage}[t]{0.5\textwidth}
\vspace*{-1ex}%
\begin{align*}
  \pi \circ \Flat &= \pi \circ \bbT\pi\,, \\
  \pi \circ \sing &= \id\,.
\end{align*}
The first of these equations is called the \emph{associative law} for~$\pi$,
the second one the \emph{unit law.}\strut
\end{minipage}%
\begin{minipage}[t]{0.5\textwidth}
\vspace*{0pt}%
\centering
\includegraphics{Algebras-submitted-1.mps}\par
%
%
%
%
%
\end{minipage}

A pair $\langle A,\pi\rangle$ satisfying these two laws is called a \emph{$\bbT$-algebra.}
For such a $\bbT$-algebra to be well-behaved, the operations $\bbT$, $\Flat$, and $\sing$
should harmonise with each other. As it turns out, three equations are sufficient.
\begin{Def}
Let $\calC$~be a category.
A triple $\langle \bbT, \Flat, \sing\rangle$ consisting of a functor
$\bbT : \calC \to \calC$ and two natural transformations
$\Flat : \bbT \circ \bbT \Rightarrow \bbT$ and $\sing : \mathrm{Id} \Rightarrow \bbT$
is a \emph{monad} if
\begin{align*}
  \Flat \circ \sing = \mathrm{id}\,, \qquad
  \Flat \circ \bbT\sing = \mathrm{id}\,, \qquad
  \Flat \circ \mu = \Flat \circ \bbT\Flat\,.
\end{align*}

\medskip
{\centering
\includegraphics{Algebras-submitted-2.mps}\par
%
%
%
%
\par}

\vskip-1.2em

\noindent
\markenddef
\end{Def}

Note that the first and third equation above are just the associative and unit laws
for the algebra $\langle\bbT X,\Flat\rangle$. This algebra is called the
\emph{free algebra} generated by~$X$.

In our framework, we will adopt this setting of monads and $\bbT$-algebras.
We will use a functor~$\bbT$ mapping a set~$A$ to the set of all $A$-labelled trees,
and a flattening operation $\Flat : \bbT\bbT A \to \bbT A$ that takes a tree labelled
by small trees and assembles these into a single large one.
Before giving the precise definitions, we need to set up a few preliminaries.

First, we us introduce the category we will be working in.
As we have chosen to work with ranked trees and we will be working with ordered
algebras, we use the category of ordered and ranked sets.
\begin{Def}
(a) A \emph{ranked set} is a sequence $A = (A_n)_{n<\omega}$ of sets~$A_n$.
The members of~$A_n$ are called \emph{elements of arity}~$n$.
We will tacitly identify such a sequence with its disjoint union $A = \bigcupdot_n A_n$.
This union is equipped with an \emph{arity function} $\ar : A \to \omega$
mapping every element $a \in A_n$ to its arity~$n$.

(b) An \emph{ordered set} $\langle A,{\leq}\rangle$ consists of a ranked set~$A$ and
a partial order~$\leq$ on~$A$ such that elements of different arities are incomparable.
(Equivalently, we can consider~$\leq$ as a sequence $({\leq_n})_n$ where
$\leq_n$~is a partial order on~$A_n$.)
Usually, we will omit the order~$\leq$ from the notation and denote an ordered set just by
its domain~$A$.
\markenddef
\end{Def}

\begin{Def}
Let $A$~and~$B$ be ordered sets and $f : A \to B$ a partial function.

(a) The \emph{domain} of~$f$ is the set
\begin{align*}
  \dom(f) := \set{ a \in A }{ f(a) \text{ is defined} }\,.
\end{align*}

(b) $f$~is \emph{monotone} if
\begin{align*}
  a \leq b
  \qtextq{implies}
  f(a) \leq f(b)\,,
  \quad\text{for all } a,b \in \dom(f)\,.
\end{align*}

(c) $f$~is a \emph{partial function of ordered sets} if it is monotone and
it preserves arities.
If $f$~is total, we call it a \emph{(total) function of ordered sets.}

\medskip
(d) $\pPos$ denotes the category of all partial functions of ordered sets
and $\Pos$ the subcategory of all total ones.
\markenddef
\end{Def}

\begin{Rem}
The categories $\pPos$ and $\Pos$ are complete and cocomplete, that is,
they have all small limits and colimits.
For instance, in the category~$\Pos$ the \emph{product} $A \times B$ of two ordered sets
$A$~and~$B$ is given by
\begin{align*}
  (A \times B)_n = A_n \times B_n
\end{align*}
with the component-wise ordering.
The \emph{coproduct} $A+B$ is given by the disjoint union
\begin{align*}
  (A + B)_n = A_n + B_n
\end{align*}
where the ordering is induced by those of $A$~and~$B$
with elements from different sets being incomparable.
\end{Rem}

Our functor~$\bbT$ will map a ranked set~$A$ to the set of all ranked trees
labelled by elements from~$A$. In addition, we will allow leaves of such trees
to be labelled with variables $x_0,x_1,x_2,\dots$ instead.
Let us start by defining what we mean by a tree.
\begin{Def}
Let $D$~be a set (unranked).

(a) We denote by $D^{<\omega}$ the set of all finite sequences of elements
of~$D$. $D^\omega$~is the set of all infinite sequences and
$D^{\leq\omega} := D^{<\omega} \cup D^\omega$.
The \emph{empty sequence} is~$\emptyseq$.

(b) The \emph{prefix ordering} on~$D^{\leq\omega}$ is
\begin{align*}
  x \preceq y \quad\defiff\quad y = xz
  \quad\text{for some } z \in D^{\leq\omega}.
\end{align*}
If $y = xd$ for $x \in D^{<\omega}$ and $d \in D$,
we say that $y$~is an \emph{(immediate) successor} of~$x$
and $x$~is an \emph{(immediate) predecessor} of~$y$.

(c)
Let $w \in D^{\leq\omega}$.
The \emph{length} $\abs{w}$ of~$w$
is the ordinal $\alpha \leq \omega$ such that $w \in D^\alpha$.
We write $w \restriction n$ for the prefix of~$w$ of length~$n$
and we denote the elements of the sequence~$w$ by $w_n$ or by $w(n)$,
for $n < \abs{w}$.
Thus, $w \restriction n+1 = (w \restriction n)w_n$.
\markenddef
\end{Def}

\begin{Def}
Let $A$~be a ranked set.

(a) A \emph{tree domain} is a non-empty set $D \subseteq \omega^{<\omega}$
such that, for all $u \in \omega^{<\omega}$ and $k < \omega$,
\begin{itemize*}
\item $u \in D$ implies $v \in D$, for all $v \prec u$,
\item $uk \in D$ implies $ui \in D$, for all $i < k$.
\end{itemize*}

(b) An \emph{$A$-labelled tree} is a function $t : \dom(t) \to A$
where $\dom(t) \subseteq \omega^{<\omega}$ is a tree domain
and every vertex $v \in \dom(t)$ has exactly $\ar(t(v))$
immediate successors.
We call the number $\ar(v) := \ar(t(v))$ the \emph{arity} of the vertex~$v$.

(c) A \emph{branch} of a tree~$t$ is a sequence
$\beta \in \omega^{\leq\omega}$ such that
$\beta \restriction n \in \dom(t)$, for all finite $n \leq \abs{\beta}$,
and $\dom(t)$ contains no successor of~$\beta$.
Hence, a branch~$\beta$ is either finite and $\beta \in \dom(t)$ is a leaf
of~$t$, or it is infinite and every proper prefix of~$\beta$ belongs to $\dom(t)$.
\markenddef
\end{Def}

These preliminaries out of the way we can finally define our functor~$\bbT$.
\begin{Def}
Let $A$~be an ordered set.
For $n < \omega$, we denote by $\bbT_n A$
the set of all $(A \cupdot \{x_0,\dots,x_{n-1}\})$-labelled trees~$t$
where $x_0,\dots,x_{n-1}$ are new $0$-ary symbols and, for every $i < n$,
there is at most one vertex $v \in \dom(t)$ labelled by~$x_i$
and this vertex is not the root of~$t$.
The union is $\bbT A := \bigcupdot_n \bbT_n A$.

Vertices labelled by a variable~$x_i$ are called \emph{holes,} or \emph{ports,}
with label~$i$.
For $t \in \bbT_n A$, we denote the unique vertex $v \in \dom(t)$
with label~$x_i$ by $\hole_i(t)$. If there is no such vertex, we leave
$\hole_i(t)$ undefined. The set of all holes is
\begin{align*}
  \Hole(t) := \set{ v \in \dom(t) }{ t(v) \in \{x_0,\dots,x_{n-1}\} }\,.
\end{align*}
\upqed
\markenddef
\end{Def}

To make $\bbT$~into a functor we also have to define the ordering on~$\bbT A$
and we have to define the operation of~$\bbT$ on functions.
The ordering is defined component-wise and $\bbT f$ applies the function~$f$
to all labels. The formal definitions are as follows.
\begin{Def}
(a) For a partial function $f : A \to B$ of ordered sets, we denote by
\begin{align*}
  \bbT f : \bbT A \to \bbT B
\end{align*}
the function that, given a  tree $t \in \bbT_n A$,
returns the tree $t' \in \bbT_n B$ obtained from~$t$ by applying~$f$
to each label, that is,
$\dom(t') = \dom(t)$ and
\begin{align*}
  t'(v) = \begin{cases}
            f(t(v)) &\text{if } t(v) \in A\,, \\
            t(v)    &\text{if } v \in \Hole(t)\,,
          \end{cases}
  \quad\text{for all } v \in \dom(t)\,.
\end{align*}
We let $\bbT f(t)$~be undefined,
if there is some vertex $v \notin \Hole(t)$ such that $f(t(v))$ is undefined.

(b) Two trees $s \in \bbT A$ and $t \in \bbT B$ have \emph{the same shape}
if they have the same domains and the same holes (with the same numbering).
We denote this relation by $s \simeq_\sh t$.
We can formally define it by setting
\begin{align*}
  s \simeq_\sh t \quad\defiff\quad
  &\text{there is some } u \in \bbT C \text{ and functions }
  p : C \to A\,,\ q : C \to B \\
  &\text{such that }
  \bbT p(u) = s \text{ and } \bbT q(u) = t\,.
\end{align*}

(c) For a binary relation $\theta \subseteq A \times B$ and
two trees $s \in \bbT A$ and $t \in \bbT B$, we write
\begin{align*}
  s \mathrel\theta^\bbT t \quad\defiff\quad
  s \simeq_\sh t \!\qtextq{and}\!
  s(v) \mathrel\theta t(v)
  \!\quad\text{for all } v \in \dom(s) \setminus \Hole(s)\,.\!
\end{align*}

(d) We consider~$\bbT A$ as an ordered set with order $\leq^\bbT$
where $\leq$~is the order of~$A$.
\markenddef
\end{Def}
Below we will use relations of the form~$\theta^\bbT$ mostly for
the ordering $\theta = {\leq}$ and the membership relation $\theta = {\in}$.
Thus, $\leq^\bbT$~is the componentwise ordering of two trees
and $\in^\bbT$~checks that each label of the first tree is an element of
the set labelling the corresponding vertex of the second tree.

\begin{Lem}
The operation\/~$\bbT$ is a functor\/ $\pPos \to \pPos$.
Its restriction to\/ $\Pos$ is a functor\/ $\Pos \to \Pos$.
\end{Lem}

Having found a suitable functor~$\bbT$, we next show that it forms a monad
by providing flattening and singleton functions.
\begin{Def}
Let $A$~be an ordered set and $t \in \bbT\bbT A$ a tree.

(a)
The \emph{flattening function}
\begin{align*}
  \Flat_A : \bbT \bbT A \to \bbT A
\end{align*}
maps a tree~$t$ to the tree $\Flat_A(t) : D \to A$ with domain
\begin{align*}
  D := \biglset v_0\dots v_{n-1}w \bigmset {}
          & \text{there is } z \in \dom(t) \text{ such that } \abs{z} = n\,, \\
          & w \in \dom(t(z)) \setminus \Hole(t(z)) \text{ and} \\
          & v_i = \hole_{z(i)}(t(z \restriction i)) \text{ for } i < n
            \bigrset
\end{align*}
and labelling
\begin{align*}
  \Flat_A(t)(v_0\dots v_{n-1}w) := t(z)(w)\,,
  \quad\text{for } z \in \dom(t) \text{ as above.}
\end{align*}

(b)
The \emph{singleton function}
\begin{align*}
  \sing_A : A \to \bbT A
\end{align*}
maps an element $a \in A_n$ to the tree $t \in \bbT_n A$
with domain
\begin{align*}
  \dom(t) = \{\emptyseq,\langle 0\rangle,\dots,\langle n-1\rangle\}
\end{align*}
and labelling

\noindent
\begin{minipage}[t]{6cm}
\vspace*{-\baselineskip}%
\begin{align*}
  t(v) = \begin{cases}
           a   &\text{if } v = \emptyseq\,, \\
           x_i &\text{if } v = \langle i\rangle\,.
         \end{cases}
\end{align*}
\end{minipage}%
\begin{minipage}[t]{4cm}
\vspace*{-1ex}%
\includegraphics{Algebras-submitted-3.mps}
%
%
%
%
\end{minipage}

\medskip
\leavevmode
\upqed
\markenddef
\end{Def}

\begin{Prop}
The functor\/ $\bbT : \pPos \to \pPos$ together with the natural transformations\/
$\Flat : \bbT \circ \bbT \Rightarrow \bbT$ and\/
$\sing : \mathrm{Id} \Rightarrow \bbT$ forms a monad.
Its restriction to\/ $\Pos$ also forms a monad.
\end{Prop}
\begin{proof}
We have to show that $\Flat$ and $\sing$ are natural transformations satisfying
the equations
\begin{align*}
  \Flat \circ \sing &= \mathrm{id}\,,
  \qquad
  \Flat \circ \Flat  = \Flat \circ \bbT\Flat\,. \\
  \Flat \circ \bbT\sing &= \mathrm{id}\,,
\end{align*}
{\centering
\includegraphics{Algebras-submitted-4.mps}
%
%
%
%
\par}

\medskip\noindent
Each of these equations can be established by a straightforward but
tedious calculation.
\end{proof}

After having chosen our monad~$\bbT$, we can introduce the corresponding algebras.
For technical reasons, we not only define algebras where the product function
is total, but also ones where the product is only pratially defined.
\begin{Def}
(a)
A~\emph{partial tree algebra} is a $\bbT$-algebra
where we consider $\bbT$~as a functor on~$\pPos$.
A~\emph{\textup(total\textup) tree algebra} is a $\bbT$-algebra
where we consider $\bbT$~as a functor on~$\Pos$.
We use the notation $\frakA = \langle A,\pi,{\leq}\rangle$ for (partial) tree algebras
where the ranked set~$A$ is the \emph{universe} of~$\frakA$ and $\pi$~its
\emph{product function.}

(b) A \emph{morphism $f : \frakA \to \frakB$ of partial tree algebras} is a
total function $f : A \to B$ of ordered sets that preserves the product, i.e.,

\noindent
\begin{minipage}[t]{0.55\textwidth}
\vspace*{-1ex}%
\begin{align*}
  f \circ \pi = \pi \circ \bbT f\,.
\end{align*}
\end{minipage}%
\begin{minipage}[t]{0.45\textwidth}
\vspace*{0pt}%
\centering
\includegraphics{Algebras-submitted-5.mps}\par
%
%
%
%
%
\end{minipage}

\medskip
\noindent
If $\frakA$~and~$\frakB$ are total, we call $f$~a
\emph{morphism of total tree algebras.}

We denote the category of all partial tree algebras and their morphisms by $\pAlg$,
and that of all total ones by~$\Alg$.
\markenddef
\end{Def}

To write down finite trees we will use the usual term notation.
For instance, $a(b_0,\dots,b_{n-1})$ denotes the tree with domain
$\dom(t) = \{\emptyseq,\langle 0\rangle,\dots,\langle n-1\rangle\}$ and
labelling
\begin{align*}
  t(v) = \begin{cases}
           a   &\text{if } v = \emptyseq\,, \\
           b_i &\text{if } v = \langle i\rangle\,.
         \end{cases}
\end{align*}

In the motivating example above we have said that the functor~$\bbT$ should map
a set to the free algebra generated by it.
If $\bbT$~is a monad, this is automatically the case.
\begin{Thm}\label{Thm: existence of free algebra}
For each ranked set~$X$, there exists a free tree algebra over~$X$.
It has the form $\langle\bbT X,\Flat,{\leq^\bbT}\rangle$.
\end{Thm}
\begin{proof}
The fact that $\Flat : \bbT\bbT X \to \bbT X$
is the free $\bbT$-algebra is a standard result in category theory.
As the functor~$\bbT$ is a monad, it is left adjoint to the
forgetful functor $\bbU : \Alg \to \Pos$ which maps a tree algebra~$\frakA$
to its universe~$A$ (see, e.g., Proposition~4.1.4 of~\cite{Borceux94b}).
Consequently, there exists, for every tree algebra~$\frakA$ and every function
$f : X \to A$,
a unique morphism $\varphi : \bbT X \to \frakA$ such that $f = \varphi \circ \sing$.
\end{proof}

\begin{Exam}
Let $\Sigma := \{a,b\}$ where $a$~and~$b$ are both binary symbols.
Suppose we want to use a morphism $\varphi : \bbT\Sigma \to \frakA$ to recognise
the set of all trees $t \in \bbT\Sigma$ that contain the label~$a$.
To do so, we have to remember one bit of information for every input tree~$t$\?:
whether or not $t$~contains an~$a$.
So we can attempt to define a tree algebra~$\frakA$ where for each arity~$n$
we have two elements\?: $0_n$~and~$1_n$.
Then the product of a tree $s \in \bbT_n A$ evaluates to~$1_n$ if at least one
label in~$s$ equals~$1_m$, and to~$0_n$ otherwise.

Unfortunately, matters are not quite that simple since we have to take products
of terms with variables into account.
For instance, when multiplying the tree $0_2(x_1,x_2) \in \bbT_3 A$
we cannot identify the result $a := \pi(0_2(x_1,x_2))$ with the value~$0_3$
since these two elements behave differently when multiplied\?:
\begin{align*}
  \pi(0_3(1_0,0_0,0_0)) = 1_0
  \qtextq{and}
  \pi(a(1_0,0_0,0_0)) = \pi(0_2(0_0,0_0)) = 0_0\,.
\end{align*}
That means we have to remember more information about the input tree\?:
which variables is contains. Consequently, we can use for our algebra elements of the form
$\langle b, u\rangle$
where $b \in \{0,1\}$ encodes whether the tree~$t$ in question contains an~$a$
and $u \subseteq [n]$ is the set of variables of~$t$.
The product is then defined in the natural way.
\end{Exam}

\subsection{Completeness and continuity}   

Let us take a closer look at the interactions between the ordering
of a tree algebra and its product.
In particular, we are interested in several notions of completeness and continuity.
\begin{Def}
An ordered set~$A$ is \emph{complete} if every subset $X \subseteq A_n$, $n < \omega$,
has a supremum and an infimum (w.r.t.~$\leq$).
It is \emph{distributive} if the supremum and infimum operations satisfy the
infinite distributive law\?:
\begin{align*}
  \inf_{i \in I} \sup_{k \in K_i} a_{ik}
  &= \sup_{\eta \in \prod_{i \in I} K_i} \inf_{i \in I} a_{i\eta(i)}\,, \\
  \sup_{i \in I} \inf_{k \in K_i} a_{ik}
  &= \inf_{\eta \in \prod_{i \in I} K_i} \sup_{i \in I} a_{i\eta(i)}\,.
\end{align*}
\upqed
\markenddef
\end{Def}
Below we will frequently use morphisms to transfer desirable properties from
one tree algebra to another one.
The next lemma is a simple example of this technique.
\begin{Lem}\label{Lem: surjective morphisms preserve completeness}
Let $f : A \to B$ be a surjective function of ordered sets
that preserves arbitrary joins.
\begin{enuma}
\item If $A$~is complete, then so is~$B$.
\item If $A$~is distributive and $\varphi$~preserves meets, then $B$~is also
  distributive.
\end{enuma}
\end{Lem}
\begin{proof}
(a) Let $Y \subseteq B_n$.
Setting $X := f^{-1}[Y]$ it follows that $f(\sup X) = \sup f[X] = \sup Y$ exists.
Hence, every subset of~$B$ has a supremum.
By a standard argument, this implies that every set also has an infimum.
(The infimum of a set is the supremum of its lower bounds.)

(b) Consider elements $b_{ik} \in B$, for $i \in I$ and $k \in K_i$.
As $\varphi$~is surjective, there are elements $a_{ik} \in \varphi^{-1}(b_{ik})$.
Consequently,
\begin{align*}
  \inf_{i \in I} \sup_{k \in K_i} b_{ik}
  &= \inf_{i \in I} \sup_{k \in K_i} \varphi(a_{ik}) \\
  &= \varphi\Bigl(\inf_{i \in I} \sup_{k \in K_i} a_{ik}\Bigr) \\
  &= \varphi\Bigl(\sup_{\eta \in \prod_{i \in I} K_i} \inf_{i \in I} a_{i\eta(i)}\Bigr) \\
  &= \sup_{\eta \in \prod_{i \in I} K_i} \inf_{i \in I} \varphi(a_{i\eta(i)})
   = \sup_{\eta \in \prod_{i \in I} K_i} \inf_{i \in I} b_{i\eta(i)}\,.
\end{align*}
\upqed
\end{proof}

We introduce two notions of continuity\?: one based on joins and one on meets.
For the latter one, we also need a restricted version, where we require
continuity only for trees labelled by a given subset of the domain.
The two definitions are not entirely symmetric since we are dealing with
partial algebras and we want to interpret an undefined result as
the least element.
\begin{Def}
Let $\frakA = \langle A,\pi,{\leq}\rangle$ be a partial tree algebra.

(a) $\frakA$~is \emph{join-continuous} if we have
\begin{align*}
  \pi(t) =
    \sup {\bigset{ \pi(s) }
                 { s \in^\bbT S \text{ and } \pi(s) \text{ is defined} }}\,,
\end{align*}
for all trees $t \in \bbT A$ and $S \in \bbT\PSet(A)$ such that
\begin{align*}
  \pi(t) \text{ is defined} \qtextq{and} t = \bbT\sup(S)\,.
\end{align*}

(b) A set $C \subseteq A$ is \emph{meet-continuously embedded} in~$\frakA$ if,
for all $S \in \bbT\PSet(C)$,
\begin{align*}
  \pi(\bbT\inf(S)) = \inf {\set{ \pi(s) }{ s \in^\bbT S }}\,,
\end{align*}
where we require both sides of this equation to be defined for the same trees~$S$
and we consider the right-hand side to be defined if every product~$\pi(s)$ is defined
and the set of these values does have a infimum.
The algebra~$\frakA$ is \emph{meet-continuous} if
its universe~$A$ is meet-con\-tinu\-ously embedded in~$\frakA$.

(c) We denote by $\CAlg$ the subcategory of $\Alg$ consisting of all
complete, distributive, and join-continuous tree algebras and all morphisms between such
algebras that preserve arbitrary joins.
\markenddef
\end{Def}

Again we collect a few technical lemmas that allow us to transfer continuity from
one algebra to another.
\begin{Lem}\label{Lem: surjective morphisms preserve join-continuity}
Let $\varphi : \frakA \to \frakB$ be a surjective morphism of tree algebras
that preserves arbitrary joins.
If\/ $\frakA$~is complete, distributive, and join-continuous, then so is\/~$\frakB$.
\end{Lem}
\begin{proof}
We have already seen in
Lemma~\ref{Lem: surjective morphisms preserve completeness}
that the algebra~$\frakB$ is complete and distributive.
For join-con\-tinu\-ity, consider trees
$t \in \bbT B$ and $S \in \bbT\PSet(B)$ with $t = \bbT\sup(S)$.
We choose some tree $S' \in \bbT\PSet(A)$ with $S = \bbT\varphi(S')$.
Setting $t' = \bbT\sup(S')$, we obtain
\begin{align*}
  t(v) = \sup S(v)
       = \sup \varphi[S'(v)]
       = \varphi(\sup S'(v))
       = \varphi(t'(v))\,,
\end{align*}
for all $v \in \dom(t)$.
Therefore,
\begin{align*}
  \pi(t)
   = \pi(\bbT\varphi(t'))
  &= \varphi(\pi(t')) \\
  &= \varphi\bigl(\sup {\set{ \pi(s') }{ s' \in^\bbT S' }}\bigr) \\
  &= \sup {\bigset{ \varphi(\pi(s')) }{ s' \in^\bbT S' }} \\
  &= \sup {\bigset{ \pi(\bbT\varphi(s')) }{ s' \in^\bbT S' }}
   = \sup {\set{ \pi(s) }{ s \in^\bbT S }}\,.
\end{align*}
\end{proof}

\begin{Lem}\label{Lem: morphisms preserving meet-continuity}
Let $\varphi : \frakA \to \frakB$ be a morphism of complete tree algebras
that preserves arbitrary meets.
If $C \subseteq A$ is meet-continuously embedded in\/~$\frakA$,
then $\varphi[C]$~is meet-continuously embedded in\/~$\frakB$.
\end{Lem}
\begin{proof}
Consider trees $t \in \bbT B$ and $S \in \bbT\PSet(\varphi[C])$ with $t = \bbT\inf(S)$.
We choose some tree $S' \in \bbT\PSet(A)$ with $S = \bbT\varphi(S')$.
Setting $t' = \bbT\inf(S')$, it follows as in the proof of
Lemma~\ref{Lem: surjective morphisms preserve join-continuity} that
\begin{align*}
  t(v) = \varphi(t'(v))
  \qtextq{and}
  \pi(t) = \inf {\set{ \pi(s) }{ s \in^\bbT S }}\,.
\end{align*}
\upqed
\end{proof}

\subsection{Join-generators}   

Below we will mostly consider tree algebras that are complete and
join-con\-tinu\-ous. Many properties of such algebras can be reduced to
corresponding properties of a subalgebra whose elements generate the
full algebra via joins.
\begin{Def}
Let $A$ be an ordered set.

(a) For a subset $S \subseteq A$, we set
\begin{align*}
  \Belowseg S &:= \set{ a \in A }{ a \leq s \text{ for some } s \in S }\,, \\
  \Aboveseg S &:= \set{ a \in A }{ a \geq s \text{ for some } s \in S }\,.
\end{align*}
For singletons $S = \{s\}$, we drop the brackets and simply write $\Belowseg s$
and $\Aboveseg s$.

(b) A set $B \subseteq A$ is a set of \emph{join-generators} of~$A$ if,
for every $a \in A$, there is some set $C \subseteq B$ with $a = \sup C$.
\markenddef
\end{Def}

The next lemma summarises some basic properties of sets of join-generators.
\begin{Lem}\label{Lem: sets of join-generators}
Let $\frakA$~be a partial tree algebra and $C \subseteq A$ a set of join-generators.
\begin{enuma}\mathindent=1em%
\item $a \leq b \quad\iff\quad c \leq a \Rightarrow c \leq b \text{ for all } c \in C\,.$
\item If $\varphi,\psi : \frakA \to \frakB$ are morphisms preserving
  arbitrary joins, then
  \begin{align*}
    \varphi \restriction C = \psi \restriction C
    \qtextq{implies}
    \varphi = \psi\,.
  \end{align*}
\item If\/ $\frakA$~is join-continuous, then
  \begin{align*}
    \pi(t) =
      \sup {\bigset{ \pi(s) }
                   { s \in \bbT C\,,\ s \leq^\bbT t\,,\ \pi(s) \text{ is defined} }}\,,
    \quad\text{for all } t \in \bbT A\,.
  \end{align*}
\end{enuma}
\end{Lem}
\begin{proof}
(a) If $a \leq b$, then $c \leq a$ implies $c \leq b$, for all $c \in A$.
Conversely, suppose that $c \leq a \Rightarrow c \leq b$, for all $c \in C$.
As $C$~is a set of join generators, it follows that
\begin{align*}
  a = \sup {\set{ c \in C }{ c \leq a }} \leq \sup {\set{ c \in C }{ c \leq b }} = b\,.
\end{align*}

(b)
Consider an element $a \in A$.
Since $C$~is a set of join-generators, we have $a = \sup B$ where
$B := \set{ c \in C }{ c \leq a }$.
As $\varphi$~and~$\psi$ preserve arbitrary joins, it follows that
\begin{align*}
  \varphi(a) = \varphi(\sup B)
       = \sup \varphi[B]
       = \sup \psi[B]
       = \psi(\sup B) = \psi(a)\,.
\end{align*}

(c)
As $t(v) = \sup {\set{ c \in C }{ c \leq t(v) }}$,
the claim follows immediately by join-continuity.
\end{proof}

The next remark can be used to simplify proofs that a given morphism
preserves meets. It is sufficient to show that it preserves meets of elements
of a set of join-generators.
\begin{Lem}\label{Lem: morphism preserving meets}
Let\/ $\frakA$~and\/~$\frakB$ be complete, distributive tree algebras
and $C \subseteq A$ a set of join-generators of\/~$\frakA$.
If a morphism $\varphi : \frakA \to \frakB$ preserves meets of elements of~$C$ and
arbitrary joins, it also preserves arbitrary meets.
\end{Lem}
\begin{proof}
Let $(a_k)_{k \in K}$ be a family of elements of~$A$. We write each
$a_k = \sup_{i \in I_k} c_{ki}$ as a join of elements $c_{ki} \in C$.
By distributivity, it follows that
\begin{align*}
  \varphi\bigl(\inf_{k \in K} a_k\bigr)
  &= \varphi\bigl(\inf_{k \in K} \sup_{i \in I_k} c_{ki}\bigr) \\
  &= \varphi\bigl(\sup_{\eta \in \prod_{k \in K} I_k}
                  \inf_{k \in K} c_{k\eta(k)}\bigr) \\
  &= \sup_{\eta \in \prod_{k \in K} I_k} \inf_{k \in K} \varphi(c_{k\eta(k)}) \\
  &= \inf_{k \in K} \sup_{i \in I_k} \varphi(c_{ki})
   = \inf_{k \in K} \varphi\bigl(\sup_{i \in I_k} c_{ki}\bigr)
   = \inf_{k \in K} \varphi(a_k)\,.
\end{align*}
\upqed
\end{proof}

\subsection{Subalgebras}   

Let us take a look at how a set of join-generators can be embedded in a tree algebra.
In particular, we are interested in the case where it induces a subalgebra.
\begin{Def}
Let $\frakA = \langle A,\pi,{\leq}\rangle$ be a partial tree algebra.

(a) A partial tree algebra $\frakB = \langle B,\pi',{\leq'}\rangle$
is a \emph{partial subalgebra} of~$\frakA$ if $B \subseteq A$ and
$\pi'$ and $\leq'$
are the restrictions of, respectively, $\pi$~and~$\leq$ to the set~$B$, i.e.,
\begin{alignat*}{-1}
  & \pi'(t) = \begin{cases}
                \pi(t) &\text{if } \pi(t) \in B\,, \\
                \text{undefined} &\text{otherwise}\,,
              \end{cases}
              &&\qquad\text{for } t \in \bbT B\,, \\[1ex]
\prefixtext{and}
  & a \leq' b \quad\iff\quad a \leq b\,, &&\qquad\text{for } a,b \in B\,.
\end{alignat*}

(b) The \emph{partial subalgebra induced} by a subset $C \subseteq A$ is the
partial tree algebra~$\frakA|_C$ with domain~$C$ and product
$\pi \restriction D$ where
\begin{align*}
  D := \set{ t \in \bbT C }{ \pi(t) \text{ is defined and } \pi(t) \in C }\,.
\end{align*}

(c) The subalgebra \emph{generated} by $C \subseteq A$ is the partial subalgebra with
domain
\begin{align*}
  \langle C\rangle :=
    \set{ \pi(t) }{ t \in \bbT C\,,\ \pi(t) \text{ defined} }\,.
\end{align*}

(d) A tree algebra~$\frakA$ is \emph{finitary} if each domain~$A_m$ is
finite and there exists a finite set $S \subseteq A$ such that
$\langle S\rangle = A$.
\markenddef
\end{Def}

Unravelling the definitions we obtain the following criterion for a set inducing
a generated subalgebra.
\begin{Lem}\label{Lem: characterisation of subalgebras}
Let\/ $\frakA$~be a partial tree algebra and $C \subseteq A$ a set.
The following statements are equivalent\?:
\begin{enum1}
\item $\langle C\rangle = C$
\item The inclusion map $i : C \to A$ is a morphism of partial tree algebras.
\item $\pi(t) \in C\,, \quad\text{for every } t \in \bbT C \text{ such that }
  \pi(t) \text{ is defined.}$
\end{enum1}
\end{Lem}

\begin{Lem}
Let $\varphi : \frakA \to \frakB$ be a morphism of partial tree algebras and
$C \subseteq A$ a set. Then
\begin{align*}
  \varphi[\langle C\rangle] = \langle\varphi[C]\rangle\,.
\end{align*}
\end{Lem}
\begin{proof}
We have
\begin{align*}
  \varphi[\langle C\rangle]
  &= \varphi\bigl[\set{ \pi(t) }{ t \in \bbT C\,,\ \pi(t) \text{ defined} }\bigr] \\
  &= \set{ \varphi(\pi(t)) }{ t \in \bbT C\,,\ \pi(t) \text{ defined} } \\
  &= \set{ \pi(\bbT\varphi(t)) }
         { t \in \bbT C\,,\ \pi(\bbT\varphi(t)) \text{ defined} } \\
  &= \set{ \pi(t) }{ t \in \bbT(\varphi[C])\,,\ \pi(t) \text{ defined} } \\
  &= \langle\varphi[C]\rangle\,.
\end{align*}
\upqed
\end{proof}

\begin{Lem}\label{Lem: surjective morphisms preserve join-generators}
Let $f : \frakA \to \frakB$ be a surjective morphism of tree algebras
that preserves arbitrary joins.
If $C \subseteq A$ is a set of join-generators of\/~$\frakA$,
then $\varphi[C]$ is a set of join-generators of\/~$\frakB$.
\end{Lem}
\begin{proof}
Let $b \in B$. Since $\varphi$~is surjective,
there is some $a \in A$ with $\varphi(a) = b$. It follows that
\begin{align*}
  b &\geq \sup {\set{ d \in \varphi[C] }{ d \leq b }} \\
    &= \sup {\set{ \varphi(c) }{ c \in C,\ \varphi(c) \leq b }} \\
    &= \varphi\bigl(\sup {\set{ c \in C }{ \varphi(c) \leq \varphi(a) }}\bigr) \\
    &\geq \varphi\bigl(\sup {\set{ c \in C }{ c \leq a }}\bigr)
     = \varphi(a) = b\,.
\end{align*}
\upqed
\end{proof}

\section{Power-set algebras}   
\label{Sect: power-set}

Below we will frequently use tree algebras where the elements are subsets
of some other tree algebra.
In this section we will study a general construction producing such tree
algebras. It can be seen as a completion operation for (partial) tree algebras.

\subsection{The power-set functor}   

We start by defining the power-set functor on $\pPos$.
Below we will then lift it to a functor on $\pAlg$.
In fact, we will define two variant, one for downwards closed sets and
one for upwards closed ones.
\begin{Def}
Let $A$ be an ordered set.

(a)
The \emph{(downward) power set} $\bbD A$ of~$A$ is the ordered set with domains
\begin{align*}
  \bbD_n A := \set{ I \subseteq A_n }{ I \text{ is downwards closed} }\,,
  \quad\text{for } n < \omega\,,
\end{align*}
and ordering
\begin{align*}
  I \leq J \quad\defiff\quad I \subseteq J\,,
  \quad\text{for } I,J \in \bbD_n A\,.
\end{align*}

(b) For a partial function $f : A \to B$ of ordered sets,
we define a function $\bbD f : \bbD A \to \bbD B$ by
\begin{align*}
  \bbD f(I) := \Belowseg f[I]\,,
  \quad\text{for } I \in \bbD A\,.
\end{align*}

(c) For each set~$A$, we define a function
$\dist_A : \bbT(\PSet(A)) \to \PSet(\bbT A)$ that maps a tree of sets to a set
of trees. The formal definition is
\begin{align*}
  \dist_A(t) := \set{ s \in \bbT A }{ s \in^\bbT t }\,,
  \quad\text{for } t \in \bbT(\PSet(A))\,.
\end{align*}
\upqed
\markenddef
\end{Def}

First, let us note that it is straightforward to check that $\bbD$~forms
a monad on~$\pPos$.
\begin{Prop}
The functor $\bbD : \pPos \to \pPos$ forms a monad where the multiplication
$\union : \bbD\bbD A \to \bbD A : X \mapsto \bigcup X$ is given by taking the union and
the singleton function $\pt : A \to \bbD A : a \mapsto \Belowseg\{a\}$ is
the principal ideal operation.
\end{Prop}

\begin{Exam}
For every ordered set~$A$, there exists a partial $\bbD$-algebra
$\langle A,\sup\rangle$, where we consider the supremum function
as a partial function $\sup : \bbD A \to A$.
A~partial function $f : A \to B$ preserves arbitrary joins if, and only if,
it is a morphism $\langle A,\sup\rangle \to \langle B,\sup\rangle$
of the corresponding $\bbD$-algebras.
\end{Exam}

To show that $\bbD$~lifts to a monad on $\pAlg$, we use a standard
technique from category theory based on distributive laws.
\begin{Def}
Let $\langle\bbS,\mu,\varepsilon\rangle$ and $\langle\bbT,\nu,\eta\rangle$ be monads.
A natural transformation $\lambda : \bbS\bbT \Rightarrow \bbT\bbS$ is a
\emph{distributive law} if
\begin{alignat*}{-1}
  \lambda \circ \mu &= \bbT\mu \circ \lambda \circ \bbS\lambda\,,
  &\qquad &\lambda \circ \varepsilon &= \bbT\varepsilon\,, \\
  \lambda \circ \bbS\nu &= \nu \circ \bbT\lambda \circ \lambda\,,
  &\qquad &\lambda \circ \bbS\eta &= \eta\,.
\end{alignat*}
\centering
\includegraphics{Algebras-submitted-6.mps}\par
\vskip-\baselineskip
\markenddef
\end{Def}
\begin{Lem}\label{Lem: dist distributive law}
The family\/ $\dist = (\dist_A)_A$ forms a distributive law
$\bbT\bbD \Rightarrow \bbD\bbT$.
\end{Lem}
\begin{proof}
First, note that $\dist_A$ is a well-defined function
$\bbT\bbD A \to \bbD\bbT A$ since
\begin{align*}
  s \leq^\bbT s' \in^\bbT t \qtextq{implies} s \in^\bbT t\,,
\end{align*}
by downwards closure of the sets $t(v)$.
Therefore, $\dist_A(t)$~is indeed a downwards closed set of trees.
Furthermore, $\dist_A$~is obviously monotone.

To see that $\dist$ is a natural transformation, let $f : A \to B$ be a partial
function of ordered sets. Then
\begin{align*}
  \bbD\bbT f(\dist_A(t))
  &= \bbD\bbT f\set{ s }{ s \in^\bbT t } \\
  &= \Belowseg \bigset{ \bbT f(s) }{ s \in^\bbT t } \\
  &= \Belowseg \bigset{ s }{ s \simeq_\sh t\,,\ s(v) \in f[t(v)] \text{ for all } v } \\
  &= \bigset{ s }{ s \simeq_\sh t\,,\ s(v) \in \Belowseg f[t(v)] \text{ for all } v } \\
  &= \bigset{ s }{ s \simeq_\sh t\,,\ s(v) \in \bbD f(t(v)) \text{ for all } v } \\
  &= \bigset{ s }{ s \in^\bbT \bbT\bbD f(t) }
   = \dist_B(\bbT\bbD f(t))\,.
\end{align*}

It remains to check the axioms of a distributive law.
\begin{align*}
  (\dist \circ \Flat)(t)
  &= \bigset{ s }{ s \in^\bbT \Flat(t) } \\
  &= \bigset{ \Flat(s) }
            { s \simeq_\sh t\,,\ s(v) \in^\bbT t(v) \text{ for all } v } \\
  &= \Belowseg\bigset{ \Flat(s) }
                     { s \simeq_\sh t\,,\ s(v) \in^\bbT t(v)
                       \text{ for all } v } \\
  &= \Belowseg\bigset{ \Flat(s) }{ s \in^\bbT t' }
     \quad\text{where } t'(v) = \set{ r }{ r \in^\bbT t(v) } \\
  &= \bbD\Flat \circ \set{ s }{ s \in^\bbT \bbT\dist(t) } \\
  &= \bbD\Flat \circ \dist \circ \bbT\dist\,, \displaybreak[0]\\[1ex]
\intertext{(Note that the second step above relies on the fact that each
hole appears at most once in~$t$. This is actually the only place where
we need this assumption.)}
  (\dist \circ \bbT\union)(t)
  &= \bigset{ s }{ s \in^\bbT \bbT\union(t) } \\
  &= \bigset{ s }{ s \simeq_\sh t\,,\ s(v) \in \textstyle\bigcup t(v)
                   \text{ for all } v } \\
  &= \Belowseg\bigset{ s }{ s \simeq_\sh t\,,\ s(v) \in \textstyle\bigcup t(v)
                            \text{ for all } v } \\
  &= \Belowseg\bigset{ s }{ s \simeq_\sh t\,,\ s(v) \in r(v) \text{ for some }
                            r(v) \in t(v) } \\
  &= \Belowseg\bigset{ s }{ s \in^\bbT r \text{ for some } r \in^\bbT t } \\
  &= \bigcup \Belowseg\bigset{ \set{ s }{ s \in^\bbR r } }{ r \in^\bbT t } \\
  &= \bigcup \Belowseg\bigset{ \dist(r) }{ r \in^\bbT t } \\
  &= \bigcup \bbD\dist \set{ r }{ r \in^\bbT t }
   = \union \circ \bbD\dist \circ \dist\,, \displaybreak[0]\\[1ex]
  (\dist \circ \sing)(I)
  &= \set{ s }{ s \in^\bbT \sing(I) } \\
  &= \set{ \sing(a) }{ a \in I } \\
  &= \Belowseg\set{ \sing(a) }{ a \in I }
   = \bbD\sing(I)\,, \displaybreak[0]\\[1ex]
  (\dist \circ \bbT\pt)(t)
  &= \set{ s }{ s \in^\bbT \bbT\pt(t) } \\
  &= \set{ s }{ s \simeq_\sh t\,,\ s(v) \in \Belowseg t(v) \text{ for all } v } \\
  &= \set{ s }{ s \leq^\bbT t }
   = \Belowseg\{t\}
   = \pt(t)\,.
\end{align*}
\upqed
\end{proof}

We can use distributive laws to lift a monad from the base category to
the category of algebras.
The following result can be found, e.g., in Section~9.2 of~\cite{BarrWells85}.
\begin{Thm}\label{Thm: distributive law of monads}
Let $\langle \bbS,\mu,\varepsilon\rangle$ and $\langle \bbT,\nu,\eta\rangle$ be
monads and $\lambda : \bbS\bbT \Rightarrow \bbT\bbS$ a distributive law.
\begin{enuma}
\item The composition\/ $\bbT\bbS$ forms a monad where multiplication and singleton
  operation are given by the morphisms
  \begin{align*}
    \nu \circ \bbT\bbT\mu \circ \bbT\lambda : \bbT\bbS\bbT\bbS \Rightarrow \bbT\bbS
    \qtextq{and}
    \eta \circ \varepsilon : \mathrm{Id} \Rightarrow \bbT\bbS\,.
  \end{align*}
\item One can lift\/~$\bbT$ to a functor on\/ $\bbS$-algebras
  that maps an\/ $\bbS$-algebra $\pi : \bbS A \to A$ to the\/ $\bbS$-algebra\/
  $\bbT\pi \circ \lambda : \bbS\bbT A \to \bbT A$.
\end{enuma}
\end{Thm}

Using this theorem and the distributive law $\dist$ we can lift the
functor~$\bbD$ to a functor on tree algebras.
\begin{Thm}\label{Thm: bbD lifts to functor on tree algebras}
We can lift\/ $\bbD : \pPos \to \pPos$ to a functor\/
$\pAlg \to \pAlg$ that maps a partial\/ $\bbT$-algebra
$\frakA = \langle A,\pi\rangle$ to the total\/ $\bbT$-algebra $\bbD\frakA$ with
domain $\bbD A$ and product
\begin{align*}
  \pi(t)
  &:= (\bbD\pi \circ \dist)(t) \\
  &= \bigset{ a \in A }
           { a \leq \pi(s) \text{ for some } s \in^\bbT t \text{ such that }
             \pi(s) \text{ is defined} }\,,
\end{align*}
for $t \in \bbT\bbD A$.
\end{Thm}
\begin{proof}
We have seen in Lemma~\ref{Lem: dist distributive law} that there is
a distributive law $\dist : \bbT\bbD \Rightarrow \bbD\bbT$.
Therefore, it follows by Theorem~\ref{Thm: distributive law of monads}
that we can lift~$\bbD$ to a functor on~$\bbT$-algebras mapping
$\pi : \bbT A \to A$ to the $\bbT$-algebra with product
\begin{align*}
  (\bbD\pi \circ \dist)(t)
  &= \Belowseg \bigset{ \pi(s) }{ s \in \dist(t)\,,\ \pi(s) \text{ defined} } \\
  &= \Belowseg \bigset{ \pi(s) }{ s \in^\bbT t\,,\ \pi(s) \text{ defined} } \\
  &= \bigset{ a \in A }{ a \leq \pi(s) \text{ for some } s \in^\bbT t
                         \text{ with } \pi(s) \text{ defined} }\,.
\end{align*}
\end{proof}

Using the functor~$\bbD$ we can give a concise definition of join-continuity.
\begin{Lem}\label{Lem: join-continuous iff sup is morphism}
A tree algebra $\frakA = \langle A,\pi\rangle$ is join-continuous if, and only
if, the supremum function ${\sup} : \bbD\frakA \to \frakA$ is a morphism
of partial tree algebras.
\end{Lem}
\begin{proof}
Recall that the product of the algebra $\bbD\frakA$ is given by
$\bbD\pi \circ \dist$. Hence, $\sup$~is a morphism of partial tree algebras
if, and only if,
\begin{align*}
  \pi \circ \bbT\sup = {\sup} \circ (\bbD\pi \circ \dist)\,.
\end{align*}

Furthermore, as $\sup X = \sup \Belowseg X$, it is sufficient in the definition
of join-con\-tinu\-ity, to only consider trees $S \in \bbT\bbD A$.
Thus, $\frakA$~is join-continuous if, and only if,
for every $S \in \bbT\bbD A$,
\begin{align*}
  \pi(\bbT\sup(S))
  = \sup {\bigset{ \pi(s) }
                 { s \in^\bbT S \text{ and } \pi(s) \text{ is defined} }}\,.
\end{align*}
Since
\begin{align*}
  ({\sup} \circ \bbD\pi \circ \dist)(S)
  = \sup {\bigset{ \pi(s) }
                 { s \in^\bbT S \text{ and } \pi(s) \text{ is defined} }}\,,
\end{align*}
the claim follows.
\end{proof}

All tree algebras of the form $\bbD\frakA$ are complete, distributive, and
join-con\-tinu\-ous.
\begin{Prop}\label{Prop: P(A) in CAlg}
$\bbD$~is a functor of the form\/ $\pAlg \to \CAlg$ where the join and meet in a\/
$\bbT$-algebra\/ $\bbD\frakA$ take the form
\begin{align*}
  \sup X = \bigcup X \qtextq{and}
  \inf X = \bigcap X\,, \quad\text{for } X \subseteq \bbD A\,.
\end{align*}
\end{Prop}
\begin{proof}
We start by proving that the order of $\bbD\frakA$ is complete and that the
joins and meets have the desired form. Let $X \subseteq \bbD A$.
Clearly,
\begin{align*}
  \bigcap X \subseteq I \subseteq \bigcup X\,,
  \quad\text{for all } I \in X\,.
\end{align*}
Furthermore, if
\begin{align*}
  K \subseteq I \subseteq L\,,
  \quad\text{for all } I \in X\,,
\end{align*}
then $K \subseteq \bigcap X$ and $\bigcup X \subseteq L$.
Hence, $\bigcap X$ and $\bigcup X$ are the meet and join of~$X$.

Since union and intersection satisfy the infinite distributive law,
it further follows that $\bbD\frakA$ is distributive.

Next, we check that every morphism of the form $\bbD\varphi : \bbD\frakA \to \bbD\frakB$
preserves joins. Let $X \subseteq \bbD A$.
\begin{align*}
  \textstyle
  \bbD\varphi(\bigcup X)
  &= \set{ b \in B }
         { \textstyle b \leq \varphi(a) \text{ for some } a \in \bigcup X } \\
  &= \set{ b \in B }
         { b \leq \varphi(a) \text{ for some } a \in I \text{ with } I \in X }
   = \bigcup_{I \in X} \bbD\varphi(I)\,.
\end{align*}

It remains to check join-continuity of $\bbD\frakA$.
By Lemma~\ref{Lem: join-continuous iff sup is morphism}, it is sufficient to prove
that $\sup : \bbD\bbD\frakA \to \bbD\frakA$ is a morphism of tree algebras, that is,
\begin{align*}
  (\bbD\pi \circ \dist) \circ \bbT{\sup}
  = {\sup} \circ (\bbD(\bbD\pi \circ \dist) \circ \dist)\,.
\end{align*}
Note that we have shown above that the supremum coincides with the union operation
$\union : \bbD\bbD \Rightarrow \bbD$, i.e., the multiplication of the monad~$\bbD$.
Consequently, we have
\begin{align*}
     {\sup} \circ (\bbD(\bbD\pi \circ \dist) \circ \dist)
  &= \union \circ \bbD\bbD\pi \circ \bbD\dist \circ \dist \\
  &= \bbD\pi \circ \union \circ \bbD\dist \circ \dist \\
  &= \bbD\pi \circ \dist \circ \bbT\union
   = (\bbD\pi \circ \dist) \circ \bbT{\sup}\,,
\end{align*}
where the second step follows from the fact that $\union$~is a natural transformation
and the third one from the axioms of a distributive law.
\end{proof}

\begin{Cor}\label{Cor: sup in CAlg}
If\/ $\frakA \in \CAlg$, then\/ $\sup : \bbD\frakA \to \frakA$ is a morphism of\/~$\CAlg$.
\end{Cor}
\begin{proof}
By Lemma~\ref{Lem: join-continuous iff sup is morphism},
$\sup$~is a morphism of $\pAlg$.
As $\frakA$~is complete, it is a total function.
To show that $\sup$~preserves joins, let $S \subseteq \bbD\frakA$. Then
\begin{align*}
  \textstyle
  \sup (\sup S) = \sup \bigcup S = \sup \set{ \sup X }{ X \in S }\,,
\end{align*}
as desired.
\end{proof}

According to the next proposition, the unit map $A \to \bbD A$
of the monad~$\bbD$ can be lifted to an embedding $\frakA \to \bbD\frakA$
of $\bbT$-algebras.
Hence, we can consider $\bbD\frakA$ as a kind of completion of~$\frakA$.
\begin{Def}
For a partial tree algebra~$\frakA$ we define
the \emph{canonical embedding} $\eta_\frakA : \frakA \to \bbD\frakA$ by
\begin{align*}
  \eta_\frakA(a) := \Belowseg a\,,
  \quad\text{for } a \in A\,.
\end{align*}
\upqed
\markenddef
\end{Def}
\begin{Prop}\label{Prop: canonical embedding preserves meets}
The canonical embedding $\eta_\frakA : \frakA \to \bbD\frakA$
is a morphism of partial tree algebras preserving meets.
Furthermore, the family $\eta = (\eta_\frakA)_\frakA$ is a
natural transformation $\eta : \mathrm{Id} \Rightarrow \bbD$.
\end{Prop}
\begin{proof}
When considered as a family of morphisms of $\pPos$,
the family~$\eta$ is just the singleton operation associated with the monad~$\bbD$.
In particular, it is a natural transformation $\mathrm{Id} \Rightarrow \bbD$.
Therefore, it remains to prove that each function~$\eta_\frakA$ is a
morphism of partial tree algebras that preserves meets.

We start by checking that $\eta_\frakA$~commutes with the product~$\pi$
of~$\frakA$. By Theorem~\ref{Thm: bbD lifts to functor on tree algebras},
the product of $\bbD\frakA$ is the morphism $\bbD\pi \circ \dist$.
Hence, the required equation is
\begin{align*}
  (\bbD\pi \circ \dist) \circ \bbD\eta
  = \bbD\pi \circ \eta = \eta \circ \pi\,,
\end{align*}
where the first step follows from the axioms of a distributive law
and the second one from the fact that $\eta$~is a natural transformation.

To see that $\eta_\frakA$~preserves meets, note that
\begin{align*}
  \eta_\frakA(\inf S)
  = \set{ c \in A }{ c \leq \inf S }
  = \bigcap {\set{ \Belowseg a }{ a \in S }}
  = \bigcap {\set{ \eta_\frakA(a) }{ a \in S }}\,.
\end{align*}
\end{proof}

\subsection{Extension problems}   

We consider the problem of extending a partial morphism $\frakA \to \frakB$
to a total one. If the domain of the given morphism is a set of join-generators
$C \subseteq A$ and the tree algebra~$\frakB$ is complete and join-continuous,
this poses no problem. In fact, this is equivalent to extend the morphism
$\frakC \to \frakB$ to a morphism $\bbD\frakC \to \frakB$.
\begin{Prop}\label{Prop: unique extension to P(A)}
For every morphism $\varphi : \frakA \to \frakB$ from an arbitrary partial tree
algebra\/~$\frakA$ into a complete, join-continuous tree algebra\/~$\frakB$,
the function
\begin{align*}
  \hat\varphi := {\sup} \circ \bbD\varphi
\end{align*}
is the unique morphism $\hat\varphi : \bbD\frakA \to \frakB$ of $\CAlg$
such that

\noindent
\begin{minipage}[t]{0.40\textwidth}
\vspace*{-1ex}%
\begin{align*}
  \varphi = \hat\varphi \circ \eta_\frakA\,.
\end{align*}
\end{minipage}%
\begin{minipage}[t]{0.60\textwidth}
\vspace*{0pt}%
\includegraphics{Algebras-submitted-7.mps}
%
%
%
%
%
\end{minipage}
\end{Prop}
\begin{proof}
Note that, by definition of the canonical embedding~$\eta_\frakB$, we have
\begin{align*}
  (\sup \circ \eta_\frakB)(b) = \sup \Belowseg b = b\,,
  \quad\text{for } b \in B\,.
\end{align*}
Thus ${\sup} : \bbD\frakB \to \frakB$ is a left inverse of~$\eta_\frakB$ and

\medskip
{\centering
\includegraphics{Algebras-submitted-8.mps}
%
%
%
%
%
\par}%

\vskip-\baselineskip%
\begin{align*}
  \hat\varphi \circ \eta_\frakA
  = {\sup} \circ \bbD\varphi \circ \eta_\frakA
  = {\sup} \circ \eta_\frakB \circ \varphi
  = \varphi\,.
\end{align*}

For uniqueness, suppose that $\psi : \bbD\frakA \to \frakB$ is another morphism
of $\CAlg$ satisfying $\psi \circ \eta_\frakA = \varphi$. Then
\begin{align*}
  \psi \restriction \rng \eta_\frakA
    = \varphi
    = \hat\varphi \restriction \rng \eta_\frakA\,.
\end{align*}
By Lemma~\ref{Lem: sets of join-generators}\,(b),
this implies that $\psi = \hat\varphi$.
\end{proof}

In particular, this statement holds for the free algebra.
\begin{Thm}
Let $X$~be a ranked set,\/ $\frakT$~the free algebra over~$X$,
and\/ $\frakA \in \CAlg$.
For every function $f : X \to A$, there exists a unique morphism
$\varphi : \bbD\frakT \to \frakA$ of $\CAlg$ such that
\begin{align*}
  \varphi \circ \eta_\frakT \circ \sing = f\,.
\end{align*}
\end{Thm}
\begin{proof}
The statement can be proved in exactly the same way as
Theorem~\ref{Thm: existence of free algebra} by simply
replacing the functor~$\bbT$ by $\bbD \circ \bbT$.
We give an alternative direct proof.

Since $\frakT$ is the free algebra of~$\Alg$ generated by~$X$,
there exists a unique morphism $\varphi_0 : \frakT \to \frakA$ of~$\Alg$
such that $\varphi_0 \circ \sing = f$.
By Proposition~\ref{Prop: unique extension to P(A)},
we can find a unique morphism $\varphi : \bbD\frakT \to \frakA$
such that $\varphi \circ \eta_\frakT = \varphi_0$.
Consequently,
\begin{align*}
  \varphi \circ \eta_\frakT \circ \sing
  = \varphi_0 \circ \sing = f\,.
\end{align*}

For uniqueness, suppose that $\psi : \bbD\frakT \to \frakA$ is
another such morphism. By uniqueness of~$\varphi_0$,
\begin{align*}
  \psi \circ \eta_\frakT \circ \sing = f
  \qtextq{implies}
  \psi \circ \eta_\frakT = \varphi_0\,.
\end{align*}
By uniqueness of~$\varphi$, it therefore follows that $\psi = \varphi$.
\end{proof}

Instead of extending morphisms, we can also consider the problem of extending
a partial product $\bbT B_0 \to B_0$ to a larger set $B \supseteq B_0$.
One way to do so is to use a second tree algebra~$\frakA$ and transfer its
product via a given function $A \to B$.
This is the content of the following lemma.
\begin{Lem}\label{Lem: transfer of product}
Let $\frakA = \langle A,\pi,{\leq}\rangle$ be a tree algebra and
$f : A \to B$ and $\pi' : \bbT B \to B$ functions of ordered sets such that
$f$~is surjective and
\begin{align*}
  f \circ \pi = \pi' \circ \bbT f\,.
\end{align*}
Then $\pi' : \bbT B \to B$ is a $\bbT$-algebra.
\end{Lem}
\begin{proof}
For associativity, note that
\begin{align*}
  \pi' \circ \bbT\pi' \circ \bbT\bbT f
  &= \pi' \circ \bbT f \circ \bbT\pi \\
  &= f \circ \pi \circ \bbT\pi \\
  &= f \circ \pi \circ \Flat \\
  &= \pi' \circ \bbT f \circ \Flat
   = \pi' \circ \Flat \circ \bbT\bbT f\,.
\end{align*}
As $f$~is surjective, so is $\bbT\bbT f$. Therefore, it follows that
\begin{align*}
  \pi' \circ \bbT\pi' = \pi' \circ \Flat\,.
\end{align*}

For the unit law, note that
\begin{align*}
  \pi' \circ \sing \circ f
   = \pi' \circ \bbT f \circ \sing
   = f \circ \pi \circ \sing
   = f = \id \circ f\,.
\end{align*}
By surjectivity of~$f$, it follows that $\pi' \circ \sing = \id$.
\end{proof}

We aim at extending a product $\pi_0 : \bbT C \to C$ defined on a set $C \subseteq A$
of join-generators to a join-continuous product $\pi : \bbT A \to A$.
Since the resulting function~$\pi$ has to satisfy
Lemma~\ref{Lem: sets of join-generators}\,(c) it follows that the given product~$\pi_0$
has to satisfy the following condition.
\begin{Def}
Let $C \subseteq A$ be ordered sets where $A$~is complete.
A monotone function $\pi_0 : \bbT C \to C$ satisfies the
\emph{join-extension condition} if, for all trees $S,S' \in \bbT\bbD C$,
\begin{align*}
  &\bbT\sup(S) = \bbT\sup(S') \\
\intertext{implies}
  &\sup {\set{ \pi_0(s) }{ s \in^\bbT S }} =
  \sup {\set{ \pi_0(s') }{ s' \in^\bbT S' }}\,.
\end{align*}
\upqed
\markenddef
\end{Def}

We need one more technical definition.
\begin{Def}\label{Def: embedding of ordered sets}
A partial function $f : A \to B$ of ordered sets is an
\emph{embedding of ordered sets} if it is total, injective, and it satisfies
\begin{align*}
  a \leq b \quad\iff\quad f(a) \leq f(b)\,,
  \quad\text{for all } a,b \in A\,.
\end{align*}
\upqed
\markenddef
\end{Def}
\begin{Prop}\label{Prop: extending product from join-generators}
Let $\frakC = \langle C,\pi_0,{\leq}\rangle$ be a partial tree algebra
and $\varphi : C \to A$ an embedding of ordered sets such that
$D := \rng \varphi$ is a set of join-generators of~$A$.
The image of~$\pi_0$ under~$\varphi$ satisfies the join-extension condition
if, and only if, there exists a unique join-continuous tree algebra
$\frakA = \langle A,\pi,{\leq}\rangle$
such that $\varphi : \frakC \to \frakA$ is a morphism of tree algebras.
Furthermore, in this case the product $\pi : \bbT A \to A$ takes the form
\begin{align*}
  \pi(t) = \sup {\set{ \pi(s) }{ s \in \bbT D\,,\ s \leq^\bbT t }}\,.
\end{align*}
\end{Prop}
\begin{proof}

$(\Leftarrow)$
Let $\pi_1 : \bbT D \to D$ be the image of~$\pi_0$ under~$\varphi$ and
let $\pi : \bbT A \to A$ be a join-continuous extension of~$\pi_1$.
For $S,S' \in \bbT\bbD D$ with $\bbT\sup(S) = \bbT\sup(S')$ it follows
by Lemma~\ref{Lem: sets of join-generators}\,(c) that
\begin{align*}
  \sup {\set{ \pi_1(s) }{ s \in^\bbT S }}
  &= \pi(\bbT\sup(S)) \\
  &= \pi(\bbT\sup(S'))
   = \sup {\set{ \pi_1(s') }{ s' \in^\bbT S' }}\,.
\end{align*}

$(\Rightarrow)$
We transfer the product of $\bbD\frakC$ to~$A$.
Let $\psi : A \to \bbD C$ be the function defined by
\begin{align*}
  \psi(a) := \set{ c \in C }{ \varphi(c) \leq a }\,.
\end{align*}
Furthermore, we set

\noindent
\begin{minipage}[t]{4cm}
\vspace*{-\baselineskip}%
\begin{align*}
  \hat\varphi &:= {\sup} \circ \bbD\varphi\,, \\
  \hat\pi_0 &:= \bbD\pi_0 \circ \dist\,, \\
  \pi &:= \hat\varphi \circ \hat\pi_0 \circ \bbT\psi\,, \\
  \hat\pi &:= \bbD\pi \circ \dist\,.
\end{align*}
\end{minipage}
\begin{minipage}[t]{7cm}
\vspace*{-2\baselineskip}%
\includegraphics{Algebras-submitted-9.mps}
\end{minipage}

\bigskip \noindent
Note that, by Theorem~\ref{Thm: bbD lifts to functor on tree algebras},
$\hat\pi_0 : \bbT\bbD C \to \bbD C$ and $\hat\pi : \bbT\bbD A \to \bbD A$
are the products of the corresponding power-set algebras.

Before proving that $\pi : \bbT A \to A$ is the desired product,
we first show that
\begin{align*}
    \hat\varphi \circ \hat\pi_0 \circ \bbT(\union \circ \bbD\psi)
  = \hat\varphi \circ \hat\pi_0 \circ \bbT(\psi \circ {\sup})\,.
\end{align*}
Given $t \in \bbT\bbD A$, define
\begin{align*}
  S(v)  &:= \set{ \varphi(c) }{ \varphi(c) \leq a \text{ for some } a \in t(v) }\,, \\
  S'(v) &:= \set{ \varphi(c) }{ \varphi(c) \leq \sup t(v) }\,.
\end{align*}
As $D$~is a set of join-generators, we have
\begin{align*}
  \sup S(v)
  &= \sup \bigcup \set{ D \cap \Belowseg a }{ a \in t(v) } \\
  &= \sup \bigset{ \sup (D \cap \Belowseg a) }{ a \in t(v) } \\
  &= \sup \set{ a }{ a \in t(v) }
   = \sup t(v)
   = \sup (D \cup \Belowseg t(v))
   = \sup S'(v)\,.
\end{align*}
Consequently, it follows from the join-extension condition that
\begin{align*}
     & (\hat\varphi \circ \hat\pi_0 \circ \bbT(\union \circ \bbD\psi))(t) \\
&\quad{}=
  \sup {\bigset{ \varphi(\pi_0(s)) }
               { s(v) \in \union(\bbD\psi(t(v))) \text{ for all } v }} \\
&\quad{}=
  \sup {\bigset{ \varphi(\pi_0(s)) }
               { s(v) \in S(v) \text{ for all } v }} \\
&\quad{}=
  \sup {\bigset{ \varphi(\pi_0(s')) }
               { s'(v) \in S'(v) \text{ for all } v }} \\
&\quad{}=
  \sup {\bigset{ \varphi(\pi_0(s')) }
               { s' \in \bbT C\,,\ 
                 \varphi(s'(v)) \leq \sup (t(v)) \text{ for all } v }} \\
&\quad{}=
  \sup {\bigset{ \varphi(\pi_0(s')) }
               { s'(v) \in \psi(\sup (t(v))) \text{ for all } v }} \\
&\quad{}=
  (\hat\varphi \circ \hat\pi_0 \circ \bbT(\psi \circ {\sup}))(t)\,.
\end{align*}

To prove that $\pi : \bbT A \to A$ is a $\bbT$-algebra,
we apply Lemma~\ref{Lem: transfer of product}.
Thus, we have to check that
\begin{align*}
  \pi \circ \bbT\hat\varphi = \hat\varphi \circ \hat\pi_0\,.
\end{align*}
First, note that, for $I \in \bbD C$,
\begin{align*}
  (\union \circ \bbD(\psi \circ \varphi))(I)
  &= \bigcup\Belowseg \set{ \psi(\varphi(a)) }{ a \in I } \\
  &= \bigcup\Belowseg \bigset{ \set{ c \in C }{ \varphi(c) \leq \varphi(a) } }
                             { a \in I } \\
  &= \Belowseg \bigset{ c \in C }
                      { \varphi(c) \leq \varphi(a) \text{ for some } a \in I } \\
  &= \Belowseg \bigset{ c \in C }
                      { c \leq a \text{ for some } a \in I } \\
  &= I\,.
\end{align*}
Hence, $\union \circ \bbD(\psi \circ \varphi) = \id$ and it follows that
\begin{align*}
  \pi \circ \bbT\hat\varphi
  &= \hat\varphi \circ \hat\pi_0 \circ \bbT(\psi \circ \hat\varphi) \\
  &= \hat\varphi \circ \hat\pi_0 \circ \bbT(\psi \circ {\sup} \circ \bbD\varphi) \\
  &= \hat\varphi \circ \hat\pi_0 \circ \bbT(\psi \circ {\sup}) \circ \bbT\bbD\varphi \\
  &= \hat\varphi \circ \hat\pi_0 \circ \bbT(\union \circ \bbD\psi)
       \circ \bbT\bbD\varphi \\
  &= \hat\varphi \circ \hat\pi_0 \circ \bbT(\union \circ \bbD(\psi \circ \varphi)) \\
  &= \hat\varphi \circ \hat\pi_0\,.
\end{align*}

For join-continuity, it is sufficient by
Lemma~\ref{Lem: join-continuous iff sup is morphism} to check that
$\sup : \bbD\frakA \to \frakA$ is a morphism of tree-algebras.
\begin{align*}
  {\sup} \circ \hat\pi
  &= {\sup} \circ \bbD\pi \circ \dist \\
  &= {\sup} \circ \bbD(\hat\varphi \circ \hat\pi_0 \circ \bbT\psi) \circ \dist \\
  &= {\sup} \circ \bbD({\sup} \circ \bbD\varphi) \circ \bbD(\bbD\pi_0 \circ \dist))
       \circ \bbD\bbT\psi \circ \dist \\
  &= {\sup} \circ \bbD{\sup} \circ \bbD\bbD(\varphi \circ \pi_0) \circ \bbD\dist
       \circ \dist \circ \bbT\bbD\psi \\
  &= {\sup} \circ \union \circ \bbD\bbD(\varphi \circ \pi_0) \circ \bbD\dist
       \circ \dist \circ \bbT\bbD\psi \\
  &= {\sup} \circ \bbD(\varphi \circ \pi_0) \circ \union \circ \bbD\dist
       \circ \dist \circ \bbT\bbD\psi \\
  &= \hat\varphi \circ \bbD\pi_0 \circ \dist \circ \bbT\union \circ \bbT\bbD\psi \\
  &= \hat\varphi \circ \hat\pi_0 \circ \bbT(\union \circ \bbD\psi) \\
  &= \hat\varphi \circ \hat\pi_0 \circ \bbT(\psi \circ {\sup}) \\
  &= \pi \circ \bbT{\sup}\,,
\end{align*}
where we have used the above claim,
the fact that $\sup : \bbD A \to A$ is a morphism of $\bbD$-algebras,
and that $\dist : \bbT\bbD \Rightarrow \bbD\bbT$ is a distributive law.

Finally, for uniqueness, suppose that there is another product $\pi' : \bbT A \to A$
such that $\langle A,\pi',{\leq}\rangle$ is join-continuous and
$\varphi : \frakC \to \frakA$ a morphism.
Then it follows by Lemma~\ref{Lem: sets of join-generators}\,(c) that
\begin{align*}
  \pi'(t)
  &= \sup {\bigset{ \pi'(s) }{ s \in \bbT D\,,\ s \leq^\bbT t }} \\
  &= \sup {\bigset{ \pi'(\bbT\varphi(s)) }
                  { s \in \bbT C\,,\ \bbT\varphi(s) \leq^\bbT t }} \\
  &= \sup {\bigset{ \varphi(\pi_0(s)) }
                  { s \in \bbT C\,,\ \bbT\varphi(s) \leq^\bbT t }} \\
  &= \sup {\bigset{ \pi(\bbT\varphi(s)) }
                  { s \in \bbT C\,,\ \bbT\varphi(s) \leq^\bbT t }} \\
  &= \sup {\bigset{ \pi(s) }{ s \in \bbT D\,,\ s \leq^\bbT t }} \\
  &= \pi(t)\,.
\end{align*}
\upqed
\end{proof}

\subsection{Upwards closed sets}   

If we use upwards closed sets instead of downwards closed ones, we obtain
a dual version of the power-set operation. Actually, we will slightly break
this duality by changing the behaviour of the new functor on non-total functions.
The reason for this is the fact that we would like to treat undefined values as
least elements.
\begin{Def}
Let $\frakA$ be a partial tree algebra.

(a)
The \emph{(upward) power-set algebra} $\bbU\frakA$ of~$\frakA$ has domains
\begin{align*}
  \bbU_n A := \set{ I \subseteq A_n }{ I \text{ is upwards closed} }\,,
  \quad\text{for } n < \omega\,,
\end{align*}
ordering
\begin{align*}
  I \leq J \quad\defiff\quad I \supseteq J\,,
  \quad\text{for } I,J \in \bbU_n A\,,
\end{align*}
and product
\begin{align*}
  \pi(t) := \Aboveseg\set{ \pi(s) }{ s \in^\bbT t }\,,
\end{align*}
where $\pi(t)$ remains undefined if one of the products~$\pi(s)$ is undefined.

(b) For a partial function $f : A \to B$ and a set $I \in \bbU A$, we define
a function $\bbU f : \bbU A \to \bbU B$ by
\begin{align*}
  \bbU f(I) := \Aboveseg f[I]\,,
  \quad\text{if } f(a) \text{ is defined for all } a \in I\,.
\end{align*}
Otherwise, $\bbU f(I)$ remains undefined.
\markenddef
\end{Def}

On sets the functor~$\bbU$ behaves dually to~$\bbD$ in the sense that
\begin{align*}
  \bbU A = (\bbD(A^\op))^\op, \quad\text{for } A \in \pPos\,,
\end{align*}
where ${}^\op : \pPos \to \pPos$ is the functor reversing the order
of each set. But note that the corresponding equation for functions
does not hold.
Still, using this relationship most proofs and results for~$\bbD$
transfer to~$\bbU$ with minor changes.
In the following we will therefore omit most of the proofs
and only point out the differences.

\begin{Prop}\label{Prop: bbU functor}
$\bbU : \pAlg \to \Alg$ is a functor mapping partial tree
algebras to tree algebras that are complete, distributive, and meet-continuous,
and mapping morphisms to morphisms that preserve arbitrary meets.
Join and meet of\/ $\bbU\frakA$ are given by
\begin{align*}
  \sup X = \bigcap X \qtextq{and}
  \inf X = \bigcup X\,, \quad\text{for } X \subseteq \bbU A\,,
\end{align*}
and the product is given by
\begin{align*}
  \pi(t)
  = (\bbU\pi \circ \dist)(t)
  = \bigset{ a \in A }{ a \geq \pi(s) \text{ for some } s \in^\bbT t }\,,
\end{align*}
for $t \in \bbT\bbU A$ such that $\pi(s)$ is defined for all $s \in^\bbT t$.
\end{Prop}
\begin{proof}
As above, the main part of the proof consists in showing that
$\dist$ forms a distributive law $\bbT\bbU \Rightarrow \bbU\bbT$.
Most steps in the proof of Lemma~\ref{Lem: dist distributive law}
immediately transfer to~$\bbU$.
Let us take a closer look at two parts where we need adjustments.

First, to see that $\dist$ is a natural transformation, note that
\begin{align*}
  & \bbU\bbT f(\dist_A(t)) \text{ is defined} \\
  \iff\quad
  & f(a) \text{ is defined for all } a \in t(v) \text{ and } v \in \dom(t) \\
  \iff\quad
  & \bbU f(t(v)) \text{ is defined for all } v \in \dom(t) \\
  \iff\quad
  & \bbT\bbU f(t(v)) \text{ is defined\,.}
\end{align*}
Furthermore, if these expressions are defined then
\begin{align*}
  \bbU\bbT f(\dist_A(t))
  = \bbU\bbT f(\dist_A(t))
  = \dist_B(\bbT\bbU f(t))
  = \dist_B(\bbT\bbU f(t))\,.
\end{align*}

It remains to check the axioms of a distributive law.
Note that $\bbU f(I) = (\bbD f(I^\op))^\op$, provided that $\bbU f(I)$ is defined.
Once we have shown that the expressions on both sides are defined on the same inputs,
we can therefore use duality and the equations for the functor~$\bbD$ to prove the
corresponding axioms for~$\bbU$.
Note that the only functions the functor~$\bbU$ is applied to in these axioms
are $\Flat$, $\dist$, and $\sing$, which are all total. Hence, both sides of the
equations are defined for all inputs.
\end{proof}

\begin{Lem}
Let\/ $\frakA$~be a tree algebra.
A subset $C \subseteq A$ is meet-continuously embedded in\/~$\frakA$
if, and only if, the infimum function\/ ${\inf} : \bbU C \to \frakA$ is
a morphism of partial tree algebras.
\end{Lem}
\begin{proof}
As the product of the algebra $\bbU \frakA$ is given by $\bbU\pi \circ \dist$,
it follows that $\inf$~is a morphism of partial tree algebras
if, and only if,
\begin{align*}
  \pi \circ \bbT\inf = {\inf} \circ (\bbU\pi \circ \dist)\,.
\end{align*}

Again, in the definition of meet-con\-tinu\-ity it is sufficient
to only consider trees $S \in \bbT\bbU C$.
Thus, $C$~is meet-continuously embedded in~$\frakA$ if, and only if,
for every $S \in \bbT\bbU C$,
\begin{align*}
  \pi(\bbT\inf(S)) = \inf {\set{ \pi(s) }{ s \in^\bbT S }}\,,
\end{align*}
where we use the convention that the right-hand side is defined if, and only if,
$\pi(s)$~is defined for all $s \in^\bbT S$.
Since
\begin{align*}
  ({\inf} \circ \bbU\pi \circ \dist)(S) = \inf {\set{ \pi(s) }{ s \in^\bbT S }}
\end{align*}
(with the same convention), the claim follows.
\end{proof}

\begin{Prop}
The canonical embedding
\begin{align*}
  \zeta_\frakA : \frakA \to \bbU\frakA : a \mapsto \Aboveseg a
\end{align*}
is a morphism of partial tree algebras that preserves joins.
\end{Prop}

\begin{Prop}\label{Prop: unique extension to U(A)}
For every morphism $\varphi : \frakA \to \frakB$ from an arbitrary partial tree
algebra\/~$\frakA$ into a complete, meet-continuous tree algebra~$\frakB$,
the function
\begin{align*}
  \hat\varphi := {\inf} \circ \bbU\varphi
\end{align*}
is the unique morphism $\hat\varphi : \bbU\frakA \to \frakB$
such that $\hat\varphi$~preserves meets and

\noindent
\begin{minipage}[t]{0.40\textwidth}
\vspace*{-1ex}%
\begin{align*}
  \varphi = \hat\varphi \circ \zeta_\frakA\,.
\end{align*}
\end{minipage}%
\begin{minipage}[t]{0.60\textwidth}
\vspace*{0pt}%
\includegraphics{Algebras-submitted-10.mps}
%
%
%
%
%
\end{minipage}
\end{Prop}

\begin{Def}
Let $C \subseteq A$ be ordered sets where $A$~is complete.
A monotone function $\pi_0 : \bbT C \to C$ satisfies the
\emph{meet-extension condition} if, for all trees $S,S' \in \bbT\bbU C$,
\begin{align*}
  &\bbT\inf(S) = \bbT\inf(S') \\
\intertext{implies}
  &\inf {\set{ \pi_0(s) }{ s \in^\bbT S }} =
  \inf {\set{ \pi_0(s') }{ s' \in^\bbT S' }}\,,
\end{align*}
where we again regard each side of this equation to be defined if, and only if,
the products are defined for all trees $s \in^\bbT S$ and $s' \in^\bbT S'$, respectively.
\markenddef
\end{Def}

Recall the definition of an embedding of ordered sets from
Definition~\ref{Def: embedding of ordered sets}.
\begin{Prop}\label{Prop: extending product from meet-generators}
Let\/ $\frakC = \langle C,\pi_0,{\leq}\rangle$ be a partial tree algebra,
$\varphi : C \to A$ an embedding of ordered sets, and let $B \subseteq A$
be the closure of $D := \rng \varphi$ under meets.
The image of~$\pi_0$ under~$\varphi$ satisfies the meet-extension condition
if, and only if, there exists a unique meet-continuous tree algebra\/
$\frakB = \langle B,\pi,{\leq}\rangle$
such that $\varphi : \frakC \to \frakB$ is a morphism of tree algebras.
Furthermore, in this case the product $\pi : \bbT B \to B$ takes the form
\begin{align*}
  \pi(t) = \inf {\set{ \pi(s) }{ s \in \bbT D\,,\ s \geq^\bbT t }}\,.
\end{align*}
\end{Prop}

\section{Branch-continuous algebras}   
\label{Sect: branch-continuous algebras}

\subsection{Semigroup-like algebras and traces}   

Our next aim is to develop a structure theory for tree algebras that are
generated in a certain way by an $\omega$-semigroup.
Such tree algebras will be the central notion of our framework.
In this section, we collect a bit of technical material needed for this task.
We start by noting that every tree algebra comes with canonical embeddings
$A_m \to A_n$, for $m \leq n$.
\begin{Def}
Let $\frakA$~be a tree algebra and
$\sigma : [m] \to [n]$ an injective function with $m \leq n < \omega$.
The \emph{$\sigma$-cylinder} over an element $a \in A_m$ is

\noindent
\begin{minipage}[t]{0.60\textwidth}
\vspace*{-1ex}%
\begin{align*}
  \cy_\sigma(a) := a(x_{\sigma(0)},\dots,x_{\sigma(m-1)}) \in A_n\,.
\end{align*}
\end{minipage}%
\begin{minipage}[t]{0.40\textwidth}
\vspace*{0pt}%
\includegraphics{Algebras-submitted-11.mps}
%
%
%
%
%
%
%
\end{minipage}

\medskip
\noindent
In the special case where $\ar(a) = 1$, we also use the short hand
\begin{align*}
  \cy_k(a) := a(x_k) = \cy_\sigma(a)\,,
  \quad\text{where } \sigma : [1] \to [n] : 0 \mapsto k\,.
\end{align*}
\upqed
\markenddef
\end{Def}

A further tool we will need is the unravelling operation.
To define it, we need a notion of `which variables actually appear in a label $a \in A$'.
For this reason we introduce what we call \emph{cylindrical structures.}
\begin{Def}
Let $A$~be an ordered set.

(a) A \emph{cylindrical structure} of~$A$ is a function associating
with every element $a \in A$ a pair $\langle a^0,\sigma_a\rangle$
consisting of an element $a^0 \in A$ with $\ar(a^0) \leq \ar(a)$
and a strictly increasing function $\sigma_a : [\ar(a^0)] \to [\ar(a)]$.
We require that
\begin{itemize}
\item $a \leq b \quad\Rightarrow\quad a^0 \leq b^0 \text{ and } \sigma_a = \sigma_b\,,$
\item $(a^0)^0 = a^0$ and $\sigma_{a^0} = \id$.
\end{itemize}

(b) A cylindrical structure on~$A$ is \emph{compatible} with a product
$\pi : \bbT A \to A$ if
\begin{align*}
  a = \cy_{\sigma_a}(a^0)\,,
  \quad\text{for all } a \in A\,.
\end{align*}

(c) The \emph{unravelling} of a tree $t \in \bbT A$ with respect to a given
cylindrical structure on~$A$ is the tree
\begin{align*}
  \un(t) := \Flat(S)\,,
\end{align*}
where $S \simeq_\sh t$ is defined by
\begin{align*}
  S(v) := a^0(x_{\sigma_a(0)},\dots,x_{\sigma_a(\ar(a^0)-1)}) \in \bbT_{\ar(a)} A\,,
  \quad\text{for } a := t(v)\,.
\end{align*}
\upqed
\markenddef
\end{Def}
Note that, for trees, the unravelling operation is rather simple.
It only reorders the successors of the vertices and removes unreachable subtrees.
\begin{Lem}
$\un : \bbT A \to \bbT A$ is an idempotent morphism of tree algebras.
\end{Lem}
\begin{proof}
Monotonicity of~$\un$ follows from the first condition in the definition of a cylindrical
structure, and the fact that $\un(\un(t)) = \un(t)$ from the second one.
Hence it remains to prove that $\un$~commutes with the product $\Flat$ of~$\bbT A$.
Let $t \in \bbT\bbT A$ and let $S \in \bbT\bbT\bbT A$ be a tree such that
\begin{align*}
  \un(t(v)) = \Flat(S(v))\,, \quad\text{for } v \in \dom(t)\,.
\end{align*}
This implies that $S' := \Flat(S)$ is the tree such that
\begin{align*}
  \un(\Flat(t)) = \Flat(S')\,.
\end{align*}
Consequently,
\begin{align*}
  \Flat(\bbT\un(t))
  = \Flat(\bbT\Flat(S))
  = \Flat(\Flat(S))
  = \un(\Flat(t))\,.
\end{align*}
\upqed
\end{proof}
\begin{Lem}\label{Lem: unravelling preserves product}
Let\/ $\frakA$~be a partial tree algebra whose universe~$A$ is equipped with
a cylindrical structure that is compatible with the product of\/~$\frakA$
and such that all cylinder maps~$\cy_\sigma$ are defined.
\begin{align*}
  \pi(\un(t)) = \pi(t)\,, \quad\text{for all } t \in \dom(\pi)\,.
\end{align*}
\end{Lem}
\begin{proof}
Note that (with the notation of the definition above)
\begin{align*}
  t(v) = \cy_{\sigma_{t(v)}}(t(v)^0) = \pi(S(v))
  \qtextq{implies}
  t = \bbT\pi(S)\,.
\end{align*}
Hence, $\pi(t) = \pi(\bbT\pi(S)) = \pi(\Flat(S)) = \pi(\un(t))$.
\end{proof}

\begin{Cor}
Let\/ $\frakA$~be a partial tree algebra whose universe~$A$ is equipped with
a cylindrical structure that is compatible with the product of\/~$\frakA$
and such that all cylinder maps~$\cy_\sigma$ are defined and such
that\/ $\rng \un \subseteq \dom \pi$.
The function $\hat\pi := \pi \circ \un$ is the unique total function\/ $\bbT A \to A$
that extends the product~$\pi$ of\/~$\frakA$ and such that\/
$\widehat\frakA := \langle A,\hat\pi\rangle$ is a total tree algebra.
\end{Cor}
\begin{proof}
We set $\hat\pi := \pi \circ \un$. By the preceding lemma, this is the only
possible extension of~$\pi$. To see that it in fact defines a tree algebra,
note that
\begin{align*}
  \hat\pi \circ \Flat
  = \pi \circ \un \circ \Flat
  &= \pi \circ \Flat \circ \bbT\un \\
  &= \pi \circ \bbT\pi \circ \bbT\un \\
  &= \pi \circ \bbT\hat\pi \\
  &\subseteq \pi \circ \un \circ \bbT\hat\pi
   = \hat\pi \circ \bbT\hat\pi\,,
\end{align*}
where the second but last step follows from~(a).
Since $\hat\pi$~is total, the two sides of this inclusion are equal
and $\hat\pi$~is the product of a tree algebra.
\end{proof}

Below we will be interested in ways an $\omega$-semigroup can sit inside a
tree algebra and in tree algebras generated by some $\omega$-semigroup they contain.
The basic building blocks we will use in this context are subalgebras of the following
form.
\begin{Def}
A partial tree algebra~$\frakA$ is \emph{semigroup-like} if
$A = \langle A_0 \cup A_1\rangle$.
\markenddef
\end{Def}

Note that, given a tree algebra~$\frakA$,
every subalgebra of the form~$\langle S\rangle$, for a set $S \subseteq A_0 \cup A_1$,
is semigroup-like.
In order to study semigroup-like tree algebras and to relate them to the tree algebras
they are contained in, we introduce the notion of a \emph{trace} of a tree,
which intuitively corresponds to the product of~$t$ \emph{along a single branch.}
A~trace along a given branch~$\beta$ of~$t$, is a path-shaped tree~$u$
whose labels are point-wise greater or equal to the corresponding labels of the vertices
of~$\beta$.
The formal definition is as follows.
\begin{Def}
Let $\frakA$~be a complete tree algebra, $\frakS \subseteq \frakA$ a semigroup-like
subalgebra, $t \in \bbT_m A$ a tree, and $\beta$~a branch of~$t$.

(a)
We denote by $\cl(S)$ the closure of~$S$ under non-empty meets.

(b)
An \emph{$S$-trace} of~$t$ along~$\beta$ is a tree $u \in \bbT_m(S_0 \cup S_1)$ such that
\begin{align*}
  &\dom(u) = \bigset{ 0^n }{ n < \omega,\ n \leq \abs{\beta} } \\[1ex]
  &\cy_{\beta(n)}(u(0^n)) \geq t(\beta \restriction n)\,,
  \quad\text{for all } n < \abs{\beta}\,, \\[1ex]
\prefixtext{and}
  &u(0^n) \geq t(\beta)\,, \quad\text{if } n := \abs{\beta} \text{ is finite\,.}
\end{align*}

(c)
An \emph{$S$-quasi-trace} of~$t$ along~$\beta$ is a tree $u \in \bbT_m(S \cup \{\top\})$
such that $t \leq^\bbT u$ and, for every $n < \abs{\beta}$, either
\begin{align*}
  u(\beta \restriction n) = \top
  \qtextq{or}
  u(\beta \restriction n) = \cy_{\beta(n)}(a)\,,
  \quad\text{for some } a \in S_0 \cup S_1\,.
\end{align*}
\upqed
\markenddef
\end{Def}
Note that the unravelling of a quasi-trace is a trace.
By Lemma~\ref{Lem: unravelling preserves product} it further follows that every
product in a semigroup-like tree algebra reduces to the product along some branch.
The following result collects this and a few
other characterisations of semigroup-like tree algebras.
\begin{Lem}\label{Lem: elements of semigroup-like algebras}
Let\/ $\frakA$~be a partial tree algebra such that the domain $\dom(\pi) \subseteq \bbT A$
of the product is closed under all cylinder maps\/ $\cy_\sigma : \bbT_m A \to \bbT_n A$.
The following statements are equivalent.
\begin{enum1}
\item $\frakA$ is semigroup-like.
\item Every element $a \in A$ is of the form $a = \cy_\sigma(b)$,
  for some $b \in A_0 \cup A_1$ and some injective function~$\sigma$.
\item For every tree $t \in \bbT A$ such that $\pi(t)$ is defined,
  there exists a tree $u \in \bbT(A_0 \cup A_1)$ with $\pi(u) = \pi(t)$.
\item Every tree $t \in \dom(\pi)$ has an $A$-trace~$u$ with $\pi(u) = \pi(t)$.
\end{enum1}
\end{Lem}
\begin{proof}
(2)~$\Rightarrow$~(1)
Let $a \in A$. Then $a = \cy_\sigma(b)$, for some $b \in A_0 \cup A_1$ and some~$\sigma$.
Hence, $a = \cy_\sigma(b) \in \langle A_0 \cup A_1\rangle$.

(1)~$\Rightarrow$~(3)
$\pi(t) \in A = \langle A_0 \cup A_1\rangle$ implies that $\pi(t) = \pi(u)$,
for some tree $u \in \bbT(A_0 \cup A_1)$.

(3)~$\Rightarrow$~(2)
Given an element $a \in A_n$, we can use~(3) to find a tree
$u \in \bbT_n(A_0 \cup A_1)$ such that $a = \pi(\sing(a)) = \pi(u)$.
We distinguish two cases.
If~$u$ does not contain a variable, then $u = \cy_\sigma(u')$
for some $u' \in \bbT_0(A_0 \cup A_1)$ and $\sigma : \emptyset \to [n]$.
Consequently, we have $\pi(u) = \pi(\cy_\sigma(u')) = \cy_\sigma(\pi(u'))$.
Since $\pi(u') \in A_0$, the claim follows.

If $u$~does contain a variable~$x_k$, then $u = \cy_\sigma(u')$
where $\sigma(0) = k$ and $u' \in \bbT_1(A_0 \cup A_1)$
is the tree obtained from~$u$ by replacing~$x_k$ by~$x_0$.
As above, it follows that $\pi(u) = \cy_k(\pi(u'))$ and $\pi(u') \in A_1$.

(4)~$\Rightarrow$~(3)
is trivial since an $A$-trace is an element of $\bbT(A_0 \cup A_1)$.

(2)~$\Rightarrow$~(4)
Let $u := \un(t)$ with the cylindrical structure given by~(2).
Then $u$~is an $A$-trace of~$t$ and
Lemma~\ref{Lem: unravelling preserves product} implies that $\pi(t) = \pi(u)$.
\end{proof}

In some cases, the product of a tree is determined by the products of its $S$-traces,
or its $S$-quasi-traces. We start by transforming traces into quasi-traces.
\begin{Lem}\label{Lem: traces and quasi-traces}
Let\/ $\frakA$~be a complete tree algebra, $\frakS \subseteq \frakA$ a semigroup-like
subalgebra, $t \in \bbT A$ a tree, and $\beta$~a branch of~$t$.
For every $S$-trace~$u$ of~$t$ along~$\beta$, there exists an
$S$-quasi-trace~$\hat u$ of~$t$ along~$\beta$ with $\pi(\hat u) = \pi(u)$.
\end{Lem}
\begin{proof}
Let $u$~be an $S$-trace of~$t$ along~$\beta$.
We define $\hat u \simeq_\sh t$ by
\begin{align*}
  \hat u(v) := \begin{cases}
                 \top &\text{if } v \npreceq \beta\,, \\
                 \cy_{\beta(n)}(u(0^n))
                   &\text{if } v = \beta \restriction n \prec \beta\,, \\
                 u(0^{\abs{\beta}}) &\text{if } v = \beta\,.
               \end{cases}
\end{align*}
Then $\hat u$~is an $S$-quasi-trace, $\un(\hat u) = u$, and it follows by
Lemma~\ref{Lem: unravelling preserves product} that $\pi(\hat u) = \pi(u)$.
\end{proof}

\begin{Lem}\label{Lem: comparison of trace products}
Let\/ $\frakA$~be a complete tree algebra, $\frakS \subseteq \frakA$ a semigroup-like
subalgebra, and $t \in \bbT A$.
\begin{align*}
  \pi(t)
  &\leq \inf {\bigset{ \pi(u) }{ u \text{ an $S$-quasi-trace of } t }} \\
  &\leq \inf {\bigset{ \pi(u) }{ u \text{ an $S$-trace of } t }}\,.
\end{align*}
\end{Lem}
\begin{proof}
The first inequality follows since we have $t \leq^\bbT u$, for every $S$-quasi-trace~$u$.
The second inequality follows by Lemma~\ref{Lem: traces and quasi-traces}.
\end{proof}

In the important special case of a meet-continuously embedded subalgebra,
the above inequalities become equalities.
\begin{Prop}\label{Prop: reducing products to products over branches}
Let\/ $\frakA$~be a tree algebra, $\frakS \subseteq \frakA$ a semigroup-like subalgebra
that is meet-continuously embedded in\/~$\frakA$, and set $C := \cl(S)$.
For every tree $t \in \bbT C$,
\begin{align*}
  \pi(t)
  &= \inf {\bigset{ \pi(u) }{ u \text{ an $S$-quasi-trace of } t }} \\
  &= \inf {\bigset{ \pi(u) }{ u \text{ an $S$-trace of } t }}\,.
\end{align*}
\end{Prop}
\begin{proof}
Let $t \in \bbT C$ be a tree.
We define a tree $U \in \bbT_n\PSet(\langle S\rangle)$ by
\begin{align*}
  U \simeq_\sh t \qtextq{and} U(v) := \set{ c \in \langle S\rangle }{ c \geq t(v) }\,.
\end{align*}
Since every element of~$C$ is a non-empty meet of elements of~$\langle S\rangle$,
we have $t = \bbT\inf(U)$. Hence, it follows by meet-continuous embeddedness that
\begin{align*}
  \pi(t) = \inf {\set{ \pi(s) }{ s \in^\bbT U }}\,.
\end{align*}

By Lemma~\ref{Lem: comparison of trace products},
it is therefore sufficient to show that
\begin{align*}
  \set{ \pi(s) }{ s \in^\bbT U }
  \subseteq
  \set{ \pi(u) }{ u \text{ an $S$-trace of } t }\,.
\end{align*}
Thus, consider a tree $s \in^\bbT U$.
We can use Lemma~\ref{Lem: elements of semigroup-like algebras} to find
an $S$-trace~$u$ of~$s$ such that $\pi(u) = \pi(s)$.
Since $t \leq^\bbT s$, it follows that $u$~is also an $S$-trace of~$t$.
\end{proof}

As a first application of traces, we show that the meet-closure of a semigroup-like
subalgebra is closed under products.
\begin{Prop}\label{Prop: meet closure}
Let\/ $\frakA$~be a complete tree algebra and\/ $\frakS \subseteq \frakA$ a semigroup-like
subalgebra that is meet-continuously embedded in\/~$\frakA$.
\begin{enuma}
\item $\cl(S)$ is meet-continuously embedded in\/~$\frakA$.
\item $\langle \cl(S)\rangle = \cl(S)$
\end{enuma}
\end{Prop}
\begin{proof}
(a)
Let $t \in \bbT A$ and $T \in \bbT\PSet(\cl(S))$ be trees with $t = \bbT\inf(T)$.
For each $c \in \cl(S)$, we choose a set $U_c \subseteq S$ with
$c = \inf U_c$ and we define trees $R,U \simeq_\sh T$ by
\begin{align*}
  U(v) := \set{ U_c }{ c \in T(v) }
  \qtextq{and}
  R(v) := \bigcup U(v)\,.
\end{align*}
Then $\inf R(v) = \inf T(v) = t(v)$
and it follows by meet-continuous embeddedness of~$S$ that
\begin{align*}
  \pi(t)
  &= \inf {\set{ \pi(r) }{ r \in^\bbT R }} \\
  &= \inf {\set{ \pi(r) }{ r \in^\bbT u \text{ for some } u \in^\bbT U }} \\
  &= \inf {\bigset{ \inf {\set{ \pi(r) }{ r \in^\bbT u }} }
                  { u \in^\bbT U }} \\
  &= \inf {\bigset{ \pi(\bbT\inf(u)) }{ u \in^\bbT U }}
   = \inf {\set{ \pi(s) }{ s \in^\bbT T }}\,,
\end{align*}
where the last step follows from the fact that there is a bijective
correspondence between trees $u \in^\bbT U$ and $s \in^\bbT T$.

(b) Let $t \in \bbT(\cl(S))$.
By definition of~$\cl(S)$, there is a tree
$T \in \bbT(\PSet(S) \setminus \{\emptyset\})$ such that $t = \bbT\inf(T)$.
We can use Lemma~\ref{Lem: elements of semigroup-like algebras} to find,
for every tree $s \in^\bbT T$, an $S$-trace~$\hat s$ of~$s$ such that
$\pi(\hat s) = \pi(s)$.
Furthermore, $\hat s \in \bbT S$ implies that
$\pi(s) = \pi(\hat s) \in \langle S\rangle = S$.
By meet-continuous embeddedness of $S$ in~$\frakA$, it follows that
\begin{align*}
  \pi(t) = \inf {\set{ \pi(s) }{ s \in^\bbT T }}
         = \inf {\set{ \pi(\hat s) }{ s \in^\bbT T }}
\end{align*}
is a non-empty meet of elements in~$S$. Thus, $\pi(t) \in \cl(S)$.
\end{proof}

\subsection{$\omega$-semigroups}   

Before finally defining the class of algebras we are interested in,
let us recall some facts regarding $\omega$-semigroups.
We use a definition that facilitates a comparison with tree algebras.
\begin{Def}
(a) The \emph{word functor} $\bbW : \pPos \to \pPos$ is defined by
\begin{align*}
  \bbW_0 A &:= A_1^\omega \cup A_1^{<\omega}A_0, \\
  \bbW_1 A &:= A_1^{<\omega}, \\
  \bbW_n A &:= \emptyset\,, \quad\text{for } n > 1\,.
\end{align*}

(b)
An \emph{\textup(ordered partial\textup) $\omega$-semigroup}
$\frakS = \langle S,\pi\rangle$
is a $\bbW$-algebra $\pi : \bbW S \to S$.
We use the usual notation for products in $\omega$-semigroups.
That is, for elements $a \in S_1$ and $b \in S_0 \cup S_1$,
we write $a\cdot b$, or just $ab$, instead of $a(b)$.
Similarly, we write $\prod_{i<n} a_i$ instead of $a_0(\dots (a_{n-1})\dots)$.

We denote the category of all $\omega$-semigroups by $\SGrp$.

(c) A partial $\omega$-semigroup~$\frakS$ is \emph{meet-continuous} if, for all
sequences $w \in \bbW S$ and $U \in \bbW\PSet(S)$ with $w = \bbW\inf(U)$,
we have
\begin{align*}
  \pi(w) = \inf {\set{ \pi(u) }{ u \in^\bbW U }}\,.
\end{align*}
(As usual, $\in^\bbW$~denotes the component-wise element relation and
we require that, if one side of the equation is defined, so is the other.)
\markenddef
\end{Def}

Since we can regard words as trees without branching,
the word functor~$\bbW$ is some kind of subfunctor of the tree functor~$\bbT$.
The lemma below makes this relationship precise.
\begin{Def}
Let $\langle\bbT_0,\mu_0,\varepsilon_0\rangle$ and
$\langle\bbT_1,\mu_1,\varepsilon_1\rangle$ be monads.

(a) A natural transformation $\varphi : \bbT_0 \Rightarrow \bbT_1$ is a
\emph{morphism of monads} if
\begin{align*}
  \varepsilon_1 = \varphi \circ \varepsilon_0
  \qtextq{and}
  \mu_1 \circ (\varphi \circ \bbT_0\varphi) = \varphi \circ \mu_0\,.
\end{align*}
In this case we say that $\bbT_0$~is a \emph{reduct} of~$\bbT_1$.

(b) Let $\varrho : \bbT_0 \Rightarrow \bbT_1$ be a morphism of monads
and $\frakA = \langle A,\pi,{\leq}\rangle$ a $\bbT_1$-algebra.
The \emph{$\varrho$-reduct} of~$\frakA$ is the $\bbT_0$-algebra
$\langle A,\pi \circ \varrho_A,{\leq}\rangle$.
\markenddef
\end{Def}

\begin{Lem}\label{Lem: morphism bbW to bbT}
There exists a morphism of monads $\varrho : \bbW \Rightarrow \bbT$ satisfying
\begin{align*}
  \dist \circ \varrho = \bbD\varrho \circ \dist
  \qtextq{and}
  \dist \circ \varrho = \bbU\varrho \circ \dist
\end{align*}
(depending on whether we consider $\dist$ as natural transformations\/
$\bbT\bbD \Rightarrow \bbD\bbT$ and\/ $\bbW\bbD \Rightarrow \bbD\bbW$, or as\/
$\bbT\bbU \Rightarrow \bbU\bbT$ and\/ $\bbW\bbU \Rightarrow \bbU\bbW$).
\end{Lem}
\begin{proof}
The function $\varrho_A : \bbW A \to \bbT A$ maps a word $w \in \bbW A$
to the tree $t \in \bbT A$ with domain
\begin{align*}
  \dom(t) := \set{ 0^n }{ n < \abs{w} }
\end{align*}
and labelling
\begin{align*}
  t(0^n) := w(n)\,, \quad\text{for } n < \abs{w}\,.
\end{align*}

It is straightforward to check that $\varrho$~is a morphism of monads.
For the additional equations, note that
\begin{align*}
  \dist(\varrho(U))
   = \Belowseg\set{ t }{ t \in^\bbT \varrho(U) }
   = \Belowseg\set{ \varrho(u) }{ u \in^\bbW U }
   = \bbD\varrho(\dist(U))\,,
\end{align*}
and similarly for the functor~$\bbU$.
\end{proof}

We can associate with every a tree algebra an $\omega$-semigroup as follows.
\begin{Def}
The \emph{$\omega$-semigroup $\SG(\frakA)$ associated} with a partial tree
algebra~$\frakA$ is the $\omega$-semigroup with domains $A_0$~and~$A_1$ whose
product is inherited from that of~$\frakA$.
\markenddef
\end{Def}
\begin{Lem}
$\SG : \Alg \to \SGrp$ is a functor.
\end{Lem}

Conversely, we can associate with every $\omega$-semigroup~$\frakS$
a semigroup-like tree algebra $\TA(\frakS)$ which
consists of elements of the form $a$~or~$a(x_i)$,
for an $\omega$-semigroup element~$a$ and an optional variable~$x_i$.
\begin{Def}
(a) The \emph{tree algebra $\TA(\frakS)$ associated with} a partial
$\omega$-semi\-group~$\frakS$ has domains
\begin{align*}
  \TA_n(S) := S_0 \cup (S_1 \times [n])\,,
  \quad\text{for } n < \omega\,,
\end{align*}
and the ordering
\begin{align*}
  x \leq y \quad\defiff\quad
  & x,y \in S_0 \text{ and } x \leq y \text{ in } \frakS\,, \text{ or} \\
  & x = \langle a,i\rangle,\ y = \langle b,i\rangle \text{ for }
  a \leq b \text{ and } i < n\,.
\end{align*}
We will use the more suggestive notation $a(x_k)$ for the elements of the form $\langle a,k\rangle$.

We define the product $\pi(t)$ of a tree $t \in \bbT(\TA(S))$ as the product of its unravelling
$\un(t)$ (which is a tree with a single path) in the $\omega$-semigroup~$\frakS$.
To make this precise, we need a bit of preparation.
Let $i : S \to \TA(S)$ be the natural embedding where
\begin{align*}
  i(a) = \begin{cases}
           a      &\text{if } a \in S_0\,, \\
           a(x_0) &\text{if } a \in S_1\,.
         \end{cases}
\end{align*}
We start by defining the cylinder maps $\cy_\sigma : \TA_m(S) \to \TA_n(S)$.
\begin{align*}
  \cy_\sigma(a) :=
    \begin{cases}
      a                &\text{if } a \in S_0\,, \\
      b(x_{\sigma(k)}) &\text{if } a = b(x_k) \in S_1 \times [m]\,.
    \end{cases}
\end{align*}

For the general case, consider a tree $t \in \bbT_m(\TA(S))$.
We unravel~$t$ with respect to the following cylindrical structure.
Let $a \in A_n$.
\begin{itemize}
\item If $a = b \in S_0$, we set $a^0 := b \in A_0$ and $\sigma_a : \emptyset \to [n]$.
\item If $a = b(x_k) \in S_1 \times [n]$, we set $a^0 := b(x_0) \in A_1$ and
      $\sigma_a : [1] \to [n] : 0 \mapsto k$.
\end{itemize}
Note that the unravelling $\un(t)$ is of the form
$\un(t) = \cy_\sigma(s)$ for some $s \in \rng \varrho$.
Fix the word $u \in \bbW S$ with $s = \varrho(\bbW i(u))$.
($u$~is unique since $\varrho$~and~$i$ are injective.)
We set
\begin{align*}
  \pi(t) := \cy_\sigma(i(\pi(u)))\,.
\end{align*}
{\centering
\includegraphics{Algebras-submitted-12.mps}\par
%
%
%
%
%
}

\medskip

(b) For a morphism $\varphi : \frakS \to \frakT$ of $\omega$-semigroups,
we define the function $\TA(\varphi) : \TA(\frakS) \to \TA(\frakT)$ by
\begin{align*}
  \TA(\varphi)(a) :=
  \begin{cases}
    \varphi(a)      &\text{if } a \in S_0\,, \\
    \varphi(b)(x_k) &\text{if } a = b(x_k) \in S_1 \times [n]\,.
  \end{cases}
\end{align*}
\upqed
\markenddef
\end{Def}

\begin{Prop}\label{Prop: TA(S) is a tree algebra}
Let\/ $\frakS$~be an $\omega$-semigroup.
\begin{enuma}
\item $\TA(\frakS)$ is a semigroup-like tree algebra.
\item $\frakS$~is meet-continuous if, and only if,\/ $\TA(\frakS)$ is meet-continuous.
\item $\TA : \SGrp \to \Alg$ is a functor.
\end{enuma}
\end{Prop}
\begin{proof}
(c) follows from~(a) and the definition of~$\TA$.

(a) For monotonicity, let $t \leq t'$.
Using the notation from the definition of the product (with primes where appropriate),
it follows that
\begin{align*}
  \un(t) \leq \un(t')
  \quad\Rightarrow\quad
  s \leq s'
  \quad\Rightarrow\quad
  u \leq u'\,.
\end{align*}
Consequently,
\begin{align*}
  \pi(t) = \cy_\sigma(i(\pi(u))) \leq \cy_\sigma(i(\pi(u'))) = \pi(t')\,.
\end{align*}

For the unit law, let $a \in \TA(S)$ and $t := \sing(a)$.
If $a \in S_0$, then $u = \langle a\rangle$ and
\begin{align*}
  \pi(t) = \cy_\emptyset(i(\pi(u))) = \cy_\emptyset(i(a)) = a\,.
\end{align*}
If $a = b(x_k) \in S_1 \times [m]$, then $u = \langle b\rangle$ and
\begin{align*}
  \pi(t) = \cy_k(i(\pi(u))) = \cy_k(i(b)) = b(x_k) = a\,.
\end{align*}

It remains to check associativity.
Let $t \in \bbT\bbT(\TA(S))$ and set $t' := \bbT\pi(t)$.
For every vertex $v \in \dom(t)$, we fix a word~$u_v$ and a function~$\sigma_v$
such that $\pi(t(v)) = \cy_{\sigma_v}(i(\pi(u_v)))$.
Let $\beta$~be the branch of~$t'$ corresponding to $\un(t')$
and fix $\hat u$~and~$\hat\sigma$ such that $\pi(t') = \cy_{\hat\sigma}(i(\pi(\hat u)))$.
Finally, fix $u^*$~and~$\sigma^*$ such that
$\pi(\Flat(t)) = \cy_{\sigma^*}(i(\pi(u^*)))$.
It follows that $u^*$~consists of the concatenation of the~$u_v$, for~$v$ on (a prefix of)~$\beta$.
Furthermore, each element of~$\hat u$ corresponds to the product $\pi(u_v)$ for a suitable
vertex~$v$. Since the product of an $\omega$-semigroup is associative it therefore follows
that $\pi(u^*) = \pi(\hat u)$.
This implies that
\begin{align*}
  \pi(\Flat(t)) = \cy_{\sigma^*}(i(\pi(u^*)))
                = \cy_{\hat\sigma}(i(\pi(\hat u))) = \pi(\bbT\pi(t))\,.
\end{align*}

(b) $(\Rightarrow)$
Let $t \in \bbT(\TA(S))$ and $T \in \bbT\bbU(\TA(S))$ be trees such that
$t = \bbT\inf(T)$.
Since every infimum $\inf T(v)$ is defined, it follows that either
\begin{itemize}
\item $t(v) \in S_0$ and $T(v) \subseteq S_0$, or
\item $t(v) = a(x_k)$ and $T(v) = \set{ b(x_k) }{ b \in P_v }$ for some set $P_v \subseteq S_1$.
\end{itemize}
Consequently, the unravellings of $t$~and every $s \in^\bbT T$ have the same shape and correspond
to the same path in~$t$. This implies that
\begin{align*}
  \un(t) = \bbT\inf(\un(T))\,.
\end{align*}
Let $u \in \bbW S$ and $U \in \bbW\bbU(S)$ be the words corresponding to these two unravellings.
Then $u = \bbT\inf U$ and meet-continuity of~$\frakS$ implies that
\begin{align*}
  \pi(u) = \inf {\set{ \pi(w) }{ w \in^\bbW U }}\,.
\end{align*}
Consequently,
\begin{align*}
  \pi(t) = \inf {\set{ \pi(s) }{ s \in^\bbW T }}\,.
\end{align*}

$(\Leftarrow)$
Let $u \in \bbW S$ and $U \in \bbW\bbU S$ be words with $u = \bbW\inf(U)$.
We set $t := (\varrho\circ\bbW i)(u)$ and $T := \bbU(\varrho \circ \bbW i)(U)$.
Then $\pi(t) = i(\pi(u))$ and $\pi(T) = \bbU i(\pi(U))$.
As $\TA(S)$ is meet-continuous, we furthermore have
\begin{align*}
  \pi(t) = \inf \pi(T)\,.
\end{align*}
Applying $i$~to this equation it follows that $\pi(u) = \inf \pi(U)$.
\end{proof}

Let us show that the functors\/ $\TA : \SGrp \to \Alg$ and\/ $\SG : \Alg \to \SGrp$ form an
adjunction $\TA \vdash \SG$.
\begin{Prop}\label{Prop: adjunction TA |- SG}
Let\/ $\frakS$~be an $\omega$-semigroup and\/ $\frakA$~a tree algebra.
\begin{enuma}
\item For every morphism $\varphi : \frakS \to \SG(\frakA)$ of $\omega$-semigroups,
  there exists a unique morphism $\hat\varphi : \TA(\frakS) \to \frakA$ of tree algebras
  such that $\SG(\hat\varphi) = \varphi$.
\item If $\varphi$~is surjective, then $\rng \hat\varphi = \langle A_0 \cup A_1\rangle$.
\end{enuma}
\end{Prop}
\begin{proof}
(a)
Let $a \in \TA_n(\frakS) = S_0 \cup (S_1 \times [n])$.
We define $\hat\varphi(a) \in A_n$ by
\begin{align*}
  \hat\varphi(a) :=
    \begin{cases}
      \varphi(a)        &\text{if } a \in S_0\,, \\
      (\varphi(b))(x_i) &\text{if } a = b(x_i) \in S_1 \times [n]\,.
    \end{cases}
\end{align*}
Then $\SG(\hat\varphi) = \varphi$ and $\hat\varphi$~is clearly the only possible function
with this property. Hence, it remains to prove that $\hat\varphi$~is a morphism
of tree algebras.

Let $\frakT := \TA(\frakS)$.
We start by noting that
\begin{align*}
  \pi(\bbT\varphi(u)) = \varphi(\pi(u))\,,
  \quad\text{for trees } u \in \bbT\bigl(T_0 \cup T_1\bigr)\,.
\end{align*}
For the general case, consider a tree $t \in \bbT T$ and set $t' := \bbT\varphi(t)$.
Using Lemma~\ref{Lem: elements of semigroup-like algebras},
we can find a $T$-trace~$u$ of~$t$ such that $\pi(u) = \pi(t)$.
Let $\frakB \subseteq \frakA$ be the subalgebra of~$\frakA$ generated by $A_0 \cup A_1$.
Then $\frakB$~is semigroup-like and $\rng \varphi \subseteq B$.
Hence, we can use Lemma~\ref{Lem: elements of semigroup-like algebras}
to find a $B$-trace~$v'$ of $\bbT\varphi(t)$ such that $\pi(v') = \pi(\bbT\varphi(t))$.
Fix a tree $v \in \bbT(T_0 \cup T_1)$ such that $v' = \bbT\varphi(v)$.
As $\bbT\varphi(u)$ is an $B$-trace of~$\bbT\varphi(t)'$ and $v$~is a $T$-trace of~$t$,
it follows by Lemma~\ref{Lem: comparison of trace products} that
\begin{alignat*}{-1}
  \varphi(\pi(t)) = \varphi(\pi(u))
  &= \pi(\bbT\varphi(u)) \\
  &\geq \pi(\bbT\varphi(t)) = \pi(v') = \pi(\bbT\varphi(v)) &&= \varphi(\pi(v)) \\
  &&&\geq \varphi(\pi(t))\,.
\end{alignat*}
Consequently, $\varphi(\pi(t)) = \pi(\bbT\varphi(t))$.

(b) Let $\frakT := \TA(\frakS)$.
If $\varphi$~is surjective, then $\hat\varphi[T_0] = A_0$ and $\hat\varphi[T_1] = A_1$.
Hence,
\begin{align*}
  \varphi[\langle T_0 \cup T_1\rangle]
  = \bigl\langle\varphi[T_0] \cup \varphi[T_1]\bigr\rangle
  = \langle A_0 \cup A_1\rangle\,.
\end{align*}
\upqed
\end{proof}

As an application, let us prove the following characterisation
of semigroup-like tree algebras.
\begin{Prop}
A tree algebra~$\frakA$ is semigroup-like if, and only if,
there exists a surjective morphism $\varphi : \TA(\frakS) \to \frakA$,
for some $\omega$-semigroup\/~$\frakS$.
\end{Prop}
\begin{proof}
$(\Rightarrow)$
Applying Proposition~\ref{Prop: adjunction TA |- SG} to the identity morphism
$\psi : \SG(\frakA) \to \SG(\frakA)$, we obtain a morphism
$\hat\psi : \TA(\SG(\frakA)) \to \frakA$ with
$\rng \hat\psi = \langle A_0 \cup A_1\rangle = A$.
Hence, $\hat\psi$~is surjective.

$(\Leftarrow)$
Suppose that $\varphi : \TA(\frakS) \to \frakA$ is surjective and
let $T$~be the universe of $\TA(\frakS)$.
Then
\begin{align*}
  \langle A_0 \cup A_1 \rangle
  = \bigl\langle\varphi[T_0 \cup T_1]\bigr\rangle
  = \varphi[\langle T_0 \cup T_1\rangle]
  = \varphi[T] = A\,.
\end{align*}
Hence, $\frakA$~is semigroup-like.
\end{proof}

\subsection{Skeletons and branch-continuity}   

After these preparations we are finally able to define the class of tree algebras
we are interested in.
\begin{Def}
Let $\frakA$~be a tree algebra.

(a) A semigroup-like subalgebra $\frakS \subseteq \frakA$ is a \emph{skeleton}
of~$\frakA$ if
\begin{itemize}
\item $S$~is meet-continuously embedded in~$\frakA$ and
\item $\cl(S)$ is a set of join-generators of~$\frakA$.
\end{itemize}

(b)
A tree algebra~$\frakA$ is \emph{branch-continuous} if $\frakA \in \CAlg$ and
it has a skeleton.

(c) We denote be $\BAlg$ the subcategory of $\CAlg$ consisting of all
branch-con\-tinu\-ous tree algebras and all morphisms that preserve meets and joins.
\markenddef
\end{Def}

Let us start our investigation of branch-continuous tree algebras with a
summary of how to compute products in them.
\begin{Lem}\label{Lem: product in branch-continuous algebras}
Let $\frakA$~be a branch-continuous tree algebra,\/
$\frakS \subseteq \frakA$ a skeleton of\/~$\frakA$, and $C := \cl(S)$.
\begin{alignat*}{-1}
\prefixtext{\normalfont(a)}
  \pi(t) &= \sup {\bigset{ \pi(s) }
                         { s \in \bbT C\,,\ s \leq^\bbT t }}\,,
         &&\qquad\text{for } t \in \bbT A\,, \\[1ex]
\prefixtext{\normalfont(b)}
  \pi(t) &= \inf {\bigset{ \pi(u) }{ u \text{ an $S$-trace of } t }}\,,
         &&\qquad\text{for } t \in \bbT C\,.
\end{alignat*}
\end{Lem}
\begin{proof}
(a) follows by Lemma~\ref{Lem: sets of join-generators}\,(c)\?;
and (b)~by Proposition~\ref{Prop: reducing products to products over branches}.
\end{proof}

Branch-continuity is preserved by certain morphisms.
\begin{Lem}\label{Lem: surjective morphisms preserve branch-continuity}
Let\/ $\frakA$~be a branch-continuous tree algebra and
$\varphi : \frakA \to \frakB$ a surjective morphism that preserves meets and joins.
\begin{enuma}
\item If $\frakS$~is a skeleton of\/~$\frakA$,
  then $\varphi[\frakS]$~is one of\/~$\frakB$.
\item $\frakB$~is branch-continuous.
\end{enuma}
\end{Lem}
\begin{proof}
(a)
The image $T := \varphi[S]$ is a semigroup-like subalgebra of~$\frakB$ that,
according to Lemma~\ref{Lem: morphisms preserving meet-continuity},
is meet-continuously embedded in~$\frakB$.
Since $\varphi$~is surjective and it preserves meets and joins, it further follows that
$\cl(T) = \varphi[\cl(S)]$ and that this is a set of join-generators of~$\frakB$.

(b)
According to Lemma~\ref{Lem: surjective morphisms preserve join-continuity},
$\frakB$~is complete, distributive, and join-continuous.
Furthermore, we have seen in~(a) that $\frakB$~has a skeleton.
\end{proof}

There are certain canonical branch-continuous tree algebras that are `freely'
generated by their skeleton.
We will show below that every branch-continuous tree algebra is a quotient
of an algebra of this form.
\begin{Def}
Let $\frakS$~be an $\omega$-semigroup.
The \emph{branch algebra} of~$\frakS$ is
\begin{align*}
  \Branch(\frakS) := \bbD\bbU(\TA(\frakS))\,.
\end{align*}
\upqed
\markenddef
\end{Def}

Our first aim is to show that $\Branch(\frakS)$ is branch-continuous.
\begin{Thm}\label{Thm: Branch(S) branch-continuous}
The branch algebra\/ $\Branch(\frakS)$ associated with a meet-continuous
$\omega$-semigroup\/~$\frakS$ is branch-continuous.
\end{Thm}
\begin{proof}
The fact that $\Branch(\frakS) \in \CAlg$ follows by Propositions
\ref{Prop: TA(S) is a tree algebra}~and~\ref{Prop: P(A) in CAlg}.
Hence, it remains to prove that it has a skeleton.

Let $T$~be the image of the canonical embedding $\TA(\frakS) \to \Branch(\frakS)$
and set $C := \cl(T)$.
We claim that $T$~is a skeleton of $\Branch(\frakS)$.
First, note that $\bbU T = \cl(T) = C$ since $\bbU T$~is meet-generated by~$T$
and closed under meets.
Furthermore, $\bbD\bbU T = \Branch(S)$ is the closure of $C$~under joins.
Hence, $C$~is a set of join generators of $\Branch(\frakS)$.

To conclude the proof, it remains to prove that $T$~is
meet-con\-tinu\-ously embedded in $\Branch(\frakS)$.
We have seen in
Propositions \ref{Prop: canonical embedding preserves meets}~and~\ref{Prop: bbU functor}
that the embedding $\eta : \bbU\TA(\frakS) \to \bbD\bbU\TA(\frakS)$
preserves meets and that the algebra $\bbU(\TA(\frakS))$ is meet-con\-tinu\-ous.
Hence, the image~$C$ of $\eta$ is meet-con\-tinu\-ously embedded in
$\Branch(\frakS)$.
In particular, so is $T \subseteq C$.
\end{proof}

\begin{Prop}\label{Prop: extending a morphism into the skeleton to Branch}
Let\/ $\frakA$~be a branch-continuous tree algebra with skeleton\/~$\frakS$
and let\/ $\frakU$~be a meet-continuous $\omega$-semigroup.
\begin{enuma}
\item For every morphism $\varphi : \frakU \to \SG(\frakS)$ of $\omega$-semigroups,
  there exists a unique morphism $\hat\varphi : \Branch(\frakU) \to \frakA$ of
  tree algebras such that $\SG(\hat\varphi)$ extends~$\varphi$ and
  $\hat\varphi$~preserves arbitrary joins and meets.
\item If $\varphi$~is surjective, so is~$\hat\varphi$.
\end{enuma}
\end{Prop}
\begin{proof}
(a)
Let $\frakT := \TA(\frakU)$.
By Proposition~\ref{Prop: adjunction TA |- SG}, there exists a unique morphism
$\varphi_0 : \frakT \to \frakS$ extending~$\varphi$.
By Proposition~\ref{Prop: unique extension to U(A)},
we can extend~$\varphi_0$ to a unique meet-preserving morphism
$\varphi_1 : \bbU\frakT \to \frakA$ by setting
\begin{align*}
  \varphi_1(J) := \inf \varphi_0[J]\,, \quad\text{for } J \in \bbU\frakT\,.
\end{align*}
Finally, we use Proposition~\ref{Prop: unique extension to P(A)}
to extend~$\varphi_1$ to a unique join-pre\-serving morphism
$\hat\varphi : \bbD\bbU\frakT \to \frakA$ by setting
\begin{align*}
  \hat\varphi(I) := \sup \varphi_1[I]\,, \quad\text{for } I \in \bbD\bbU\frakT\,.
\end{align*}
Note that Lemma~\ref{Lem: morphism preserving meets} implies that
$\hat\varphi$~preserves arbitrary meets.

(b)
If $\varphi$~is surjective,
so is the morphism $\varphi_0 : \TA(\frakU) \to \frakS$ from the proof of~(a), i.e.,
$\rng \varphi_0 = S$.
As $\varphi_1$~preserves arbitrary meets, its range includes the closure $C := \cl(S)$.
Similarly, the range of~$\hat\varphi$ includes the closure of~$C$ under joins,
which is all of~$A$. Thus, $\hat\varphi$~is surjective.
\end{proof}

As promised above, we can show that, conversely, every branch-continuous
tree algebra is a quotient of an algebra of the form $\Branch(\frakS)$.
\begin{Thm}
Let\/ $\frakA$~be a branch-continuous tree algebra with skeleton\/~$\frakS$.
There exists a surjective morphism $\varphi : \Branch(\SG(\frakS)) \to \frakA$
that preserves joins and meets.
\end{Thm}
\begin{proof}
Let $\psi : \SG(\frakS) \to \SG(\frakS)$ be the identity morphism.
By Proposition~\ref{Prop: extending a morphism into the skeleton to Branch},
there exists a unique extension $\hat\psi : \Branch(\SG(\frakS)) \to \frakA$
which preserves joins and meets.
\end{proof}
Combining this theorem with
Lemma~\ref{Lem: surjective morphisms preserve branch-continuity},
we obtain the following characterisation of branch-continuous tree algebras
as quotients of an algebra of the form $\Branch(\frakS)$.
\begin{Cor}
A tree algebra\/~$\frakA$ is branch-continuous if, and only if,
there exists an $\omega$-semigroup\/~$\frakS$ and a surjective morphism
$\varphi : \Branch(\frakS) \to \frakA$ that preserves joins and meets.
\end{Cor}

\subsection{Closure under products}   

Our next goal is to prove that the class of branch-continuous tree algebras
is closed under finite products.
\begin{Def}
Let $\frakA^i = \langle A^i,\pi^i,{\leq}^i\rangle$, for $i \in I$, be a family
of tree algebras. The \emph{product} $\prod_{i \in I} \frakA^i$ is the
tree algebra with domains
\begin{align*}
  \prod_{i \in I} A^i_n\,, \quad\text{for } n < \omega\,,
\end{align*}
order
\begin{align*}
  (a_i)_{i \in I} \leq (b_i)_{i \in I}
  \quad\defiff\quad a_i \leq^i b_i\,, \quad\text{for all } i \in I\,,
\end{align*}
and product
\begin{align*}
  \pi(t) := (\pi^i(\bbT p_i(t))_{i \in I}
\end{align*}
where the function $p_k : \prod_{i \in I} A^i \to A^k$ projects a sequence
$(a_i)_{i \in I}$ to its $k$-th component~$a_k$.
\markenddef
\end{Def}

\begin{Lem}\label{Lem: product of tree algebras}
Let $\frakA^i = \langle A^i,\pi^i,{\leq}^i\rangle$, for $i \in I$, be a family of
tree algebras.
\begin{enuma}
\item $\prod_{i \in I} \frakA^i$ is a tree algebra.
\item Each projection $p_k : \prod_{i \in I} \frakA^i \to \frakA^k$ is a morphism
  of tree algebras preserving arbitrary joins and meets.
\item If every $\frakA^i$~is complete, distributive, and join-continuous, then so is
  $\prod_{i \in I} \frakA^i$.
\end{enuma}
\end{Lem}
\begin{proof}
(b) It follows immediately from the definitions that $p_k$~is monotone,
that it commutes with products, and that is preserves joins and meets.

(a)
Clearly, the product~$\pi$ is monotone.
Furthermore,
\begin{align*}
  p_i(\pi(\sing(a)))
  = \pi^i(p_i(\sing(a))
  = \pi^i(\sing(p_i(a)))
  = p_i(a)\,,
\end{align*}
which implies that $\pi \circ \sing = \id$.

Hence, it remains to prove associativity.
Let $t \in \bbT_n\bbT(\prod_i A^i)$. Then
\begin{align*}
  p_i(\pi(\Flat(t)))
  &= \pi^i(\bbT p_i(\Flat(t))) \\
  &= \pi^i(\Flat(\bbT p_i(t))) \\
  &= \pi^i(\bbT\pi^i(\bbT p_i(t))) \\
  &= \pi^i(\bbT p_i(\bbT\pi(t)))
   = p_i(\pi(\bbT\pi(t)))\,,
\end{align*}
which implies that $\pi \circ \Flat = \pi \circ \bbT\pi$.

(c)
For completeness, let $X \subseteq \prod_i A^i$.
As $p_i$~commutes with joins, we have
\begin{align*}
  p_i(\sup X) = \sup p_i[X]\,,
\end{align*}
which implies that $\sup X = (\sup p_i[X])_{i \in I}$.
Similarly, it follows that $\inf X = (\inf p_i[X])_{i \in I}$.

In the same way it follows that the product is distributive.
For join-continuity, let $t = \bbT\sup(S)$. Then
\begin{align*}
  \bbT p_i(t) = \bbT p_i(\bbT\sup(S)) = \bbT\sup(\bbT p_i(S))
\end{align*}
which, by join-continuity of~$\frakA^i$, implies that
\begin{align*}
  \pi^i(\bbT p_i(t)) = \sup \set{ \pi^i(s) }{ s \in^\bbT \bbT p_i(S) }\,.
\end{align*}
Consequently,
\begin{align*}
  p_i(\pi(t))
  &= \pi^i(\bbT p_i(t)) \\
  &= \sup {\bigset{ \pi^i(s) }{ s \in^\bbT \bbT p_i(S) }} \\
  &= \sup {\bigset{ \pi^i(\bbT p_i(s)) }{ s \in^\bbT S }} \\
  &= \sup {\bigset{ p_i(\pi(s)) }{ s \in^\bbT S }}
   = p_i(\sup \set{ \pi(s) }{ s \in^\bbT S })\,.
\end{align*}
\upqed
\end{proof}

\begin{Prop}
Let $\varphi^i : \frakB \to \frakA^i$, $i \in I$, be a family of tree algebra morphisms.
There exists a unique morphism $\psi : \frakB \to \prod_{i \in I} \frakA^i$ such that
\begin{align*}
  \varphi^i = p_i \circ \psi\,,
  \quad\text{for all } i \in I\,.
\end{align*}
\end{Prop}
\begin{proof}
The function $\psi(b) := (\varphi^i(b))_{i \in I}$ has the desired properties.
\end{proof}

\begin{Thm}\label{Thm: branch-continuity preserved by products}
If\/ $\frakA$~and\/~$\frakB$ are branch-continuous tree algebras, so is
their product\/ $\frakA \times \frakB$.
\end{Thm}
\begin{proof}
We have seen in Lemma~\ref{Lem: product of tree algebras},
that $\frakA \times \frakB$ is complete, distributive, and join-continuous.
Hence, it remains to find a skeleton of $\frakA \times \frakB$.
Let $\frakS$~and~$\frakT$ be skeletons of, respectively, $\frakA$~and~$\frakB$.
We claim that $\frakS \times \frakT$ is a one of the product.

Note that we have shown in Lemma~\ref{Lem: product of tree algebras}
that the projections $p_0 : A \times B \to a$ and $p_1 : A \times B \to B$
preserve projections and arbitrary meets and joins.
Consequently, we have $\cl(S \times T) = \cl(S) \times \cl(T)$.
Furthermore, the fact that $\cl(S)$ and $\cl(T)$ are sets of join-generators
implies that so is $\cl(S) \times \cl(T)$.

It remains to show that $S \times T$ is meet-continuously embedded in
$\frakA \times \frakB$.
Let $t \in \bbT(A \times B)$ and $U \in \bbT\bbU(S \times T)$ be trees with
$t = \bbT\inf(U)$.
Applying the projection~$p_i$, we obtain
\begin{align*}
  \pi(\bbT p_i(t)) = p_i(\pi(t)) = p_i(\pi(\bbT\inf U))
  = \pi(\bbT\inf(\bbT p_i(U)))\,,
\end{align*}
which implies that
\begin{align*}
  \pi(\bbT p_i(t)) = \inf {\bigset{ \pi(s) }{ s \in^\bbT \bbT p_i(U) }}\,.
\end{align*}
It follows that $\pi(t) = \inf {\set{ \pi(s) }{ s \in^\bbT U }}$.
\end{proof}

\subsection{Regular languages and recognisability}   

Having introduced branch-continuous tree algebras we can use them
to give a characterisation of the class of regular languages.
We will use the ordered version of recognisability.
\begin{Def}
Let $\frakA$~be a tree algebra.
A subset $L \subseteq A_m$ is \emph{recognised} by a morphism
$\eta : \frakA \to \frakB$ if $L = \eta^{-1}[P]$ for some
upwards closed subset $P \subseteq B_m$.
We say that $L$~is \emph{recognised} by~$\frakB$ if it is recognised
by some morphism $\frakA \to \frakB$.
\markenddef
\end{Def}

We will show that a tree language is recognisable if, and only if, it is recognised
by a finitary, branch-continuous tree algebra.
We start by showing that recognisable languages are regular.
When taking a closer look at what it means for a recognisable language to
be regular, we arrive at the following definition, which has recently been
introduced in~\cite{BlumensathBojanczykKlinXX}.
\begin{Def}
A tree algebra~$\frakA$ is \emph{regular} if it is finitary and
there exists a finite set $C \subseteq A$ of generators such that,
for every element $a \in A$, the preimage
\begin{align*}
  \pi^{-1}(a) \cap \bbT C \text{ is a regular language.}
\end{align*}
\upqed
\markenddef
\end{Def}
It is straightforward to check that the regular tree algebras recognise
precisely the regular tree languages.
\begin{Thm}[\cite{BlumensathBojanczykKlinXX}]
A tree algebra~$\frakA$ is regular if, and only if, every language
recognised by~$\frakA$ is regular.
\end{Thm}
Of course, the definition of a regular algebra was specifically chosen to make
this theorem true. But because of its cyclic nature it does not further our
understanding of the regular tree languages. What is missing is a good
algebraic characterisation telling us how regular algebras look like.
Branch-continuous algebras \emph{do} have such a characterisation and can therefore
serve as an alternative approach to regularity. We start by showing that every
branch-continuous algebra is regular.
\begin{Prop}\label{Prop: recognisable languages are definable}
Every finitary, branch-continuous tree algebra is regular.
\end{Prop}
\begin{proof}
Let $\frakA$~be finitary and branch-continuous,
$\frakS$~a skeleton of~$\frakA$, and set $C := \cl(S)$.
Fix a finite set $B \subseteq A$ of generators.
W.l.o.g.\ we may assume that $B = A_0 \cup\dots\cup A_{k-1}$, for some $k < \omega$.
We will construct $\MSO$-formulae~$\varphi_a$ defining the languages
\begin{align*}
  \pi^{-1}(a) \cap \bbT B\,, \quad\text{for } a \in A\,.
\end{align*}

First, note that, given a tree $t \in \bbT_n B$, we can encode an $S$-trace~$u$ of~$t$
by a family $(U_c)_{c \in S}$ of unary predicates such that the union
$\bigcup_c U_c$ contains the branch corresponding to the $S$-trace~$u$
and the various predicates~$U_c$ encode its labelling.
Since, in monadic second-order logic, we can compute infinite products
in finite $\omega$-semigroups, there are formulae $\vartheta_a(\bar Z)$, for $a \in A$,
that check whether $\pi(u) = a$ when given a tree $t \in \bbT_n(B \cap C)$ and
an $S$-trace~$u$ of~$t$ that is encoded in~$\bar Z$.

For trees $t \in \bbT_n C$,
we have seen in Lemma~\ref{Lem: product in branch-continuous algebras}\,(b) that
\begin{align*}
  \pi(t) = \inf {\bigset{ \pi(u) }{ u \text{ an $S$-trace of } t }}\,.
\end{align*}
Consequently, can use the formulae $\vartheta_a(\bar Z)$ to construct
formulae~$\psi_a$ that, given a tree $t \in \bbT(B \cap C)$,
check whether $\pi(t) = a$.

Finally, according to Lemma~\ref{Lem: product in branch-continuous algebras}\,(a),
we have
\begin{align*}
  \pi(t) = \sup {\bigset{ \pi(s) }{ s \in \bbT_n C\,,\ s \leq^\bbT t }}\,,
  \quad\text{for all trees } t\,.
\end{align*}
Therefore, we can use the above formulae~$\psi_a$ to construct
formulae~$\varphi_a$, for $a \in A$,
checking whether the product of a given tree $t \in \bbT_n B$ evaluates to~$a$.
\end{proof}

It remains to prove the converse\?: given a regular language we have to find
a branch-continuous algebra recognising it.
We start by fixing our terminology regarding automata.
\begin{Def}
Let $\calA = \langle Q,\Sigma,\Delta,q_0,\Omega\rangle$ be a nondeterministic
parity automaton and set $D := \rng \Omega$.

(a) Let $t \in \bbT_n\Sigma$.
A \emph{run} of~$\calA$ on a tree $t \in \bbT_n \Sigma$ is a
tree $\varrho \in \bbT_0 Q$ with the same domain as~$t$
that satisfies the following two conditions\?:
\begin{itemize}
\item for every vertex $v \in \dom(t) \setminus \Hole(t)$ with $\ar(t(v)) = n$
  and immediate successors $u_0,\dots,u_{n-1}$,
  \begin{align*}
    \bigl\langle \varrho(v),t(v),\varrho(u_0),\dots,\varrho(u_{n-1})\bigr\rangle
      \in \Delta_n\,;
  \end{align*}
\item for every infinite branch~$\beta$ of~$t$,
  \begin{align*}
    \liminf_{v \prec \beta} \Omega(\varrho(v)) \quad\text{is even.}
  \end{align*}
\end{itemize}

(b) The \emph{profile} of a run~$\varrho$ on a tree
$t \in \bbT_n\Sigma$ is the pair
\begin{align*}
  \pf(\varrho) := \bigl\langle\varrho(\emptyseq), \bar u\bigr\rangle\,,
\end{align*}
where
\begin{align*}
  u_i := \begin{cases}
           \langle d,\varrho(v_i)\rangle
             & \text{if } v_i := \hole_i(t) \text{ is defined and} \\
             &\quad d := \min {\set{ \Omega(\varrho(z)) }{ z \preceq v_i }}\,, \\
           \bot &\text{otherwise}\,.
         \end{cases}
\end{align*}
\upqed
\markenddef
\end{Def}

\begin{Def}
Let $\calA = \langle Q,\Sigma,\Delta,q_0,\Omega\rangle$ be a nondeterministic
parity automaton and set $D := \rng \Omega$.

(a)
The \emph{automaton $\omega$-semigroup}~$\frakS_\calA$ associated with~$\calA$
is the partial $\omega$-semi\-group with domains
\begin{align*}
  S_0 := Q
  \qtextq{and}
  S_1 := Q \times D \times Q\,.
\end{align*}
The order is equality on~$S_0$ and on $S_1$~it is given by
\begin{alignat*}{-1}
  \langle p,k,q\rangle \leq \langle p',k',q'\rangle
  \quad\defiff\quad &p = p',\ q = q', \text{ and }
   k \sqsubseteq k' \text{ in the ordering} \\
  &1 \sqsubset 3 \sqsubset 5 \sqsubset\dots \sqsubset 4 \sqsubset 2 \sqsubset 0\,.
\end{alignat*}
(The closer a priority is to acceptance, the larger it is.)
The product is determined by the equations
\begin{align*}
  \langle p,k,q\rangle\cdot q' &:=
    \begin{cases}
      p                &\text{if } q = q'\,,\\
      \text{undefined} &\text{otherwise}\,,
    \end{cases} \\
  \langle p,k,q\rangle\cdot\langle p',k',q'\rangle &:=
    \begin{cases}
      \langle p,l,q'\rangle &\text{if } q = p' \text{ and }
                             l := \min {\{k,k'\}}\,,\\
      \text{undefined}      &\text{otherwise}\,,
    \end{cases} \\
  \prod_{n<\omega} \langle p_n,k_n,q_n\rangle &:=
    \begin{cases}
      p_0              &\text{if } q_n = p_{n+1} \text{ for all } n
                        \text{ and} \\
                       &\liminf_{n \to \infty} k_n \text{ is even,} \\
      \text{undefined} &\text{otherwise.}
    \end{cases}
\end{align*}

(b) We define a morphism $\alpha_\calA : \bbT\Sigma \to \Branch(\frakS_\calA)$
as follows.
Given a tree $t \in \bbT_n\Sigma$, we set
\begin{align*}
  \alpha_\calA(t) :=
    \Belowseg\bigset{ \eta(\widetilde\pf(\varrho)) }{ \varrho \text{ a run on } t }\,,
\end{align*}
where $\eta : \bbU(\TA(\frakS_\calA)) \to \Branch(\frakS_\calA)$
is the canonical embedding and, for a run~$\varrho$ with profile
\begin{align*}
  \pf(\varrho) = \bigl\langle p, u_0,\dots,u_{n-1}\bigr\rangle\,,
\end{align*}
we have set
\begin{align*}
  \widetilde\pf(\varrho) :=
    \tilde p \sqcap \tilde u_0 \sqcap\dots\sqcap \tilde u_{n-1}
    \in \bbU(\TA(S_\calA))
\end{align*}
with
\begin{align*}
  \tilde u_i &:= \begin{cases}
                   \langle p,k,q\rangle(x_i) &\text{if } u_i = \langle k,q\rangle\,, \\
                   \top                      &\text{if } u_i = \bot\,,
                 \end{cases} \\
  \tilde p &:= \begin{cases}
                 p    &\text{if } t \text{ has an infinite branch or a leaf
                       that is not a hole,} \\
                 \top &\text{otherwise.}
               \end{cases}
\end{align*}
\upqed
\markenddef
\end{Def}

\begin{Lem}\label{Lem: alpha recognises L}
$\alpha_\calA : \bbT\Sigma \to \Branch(\frakS_\calA)$ is a morphism
of tree algebras recognising $L(\calA)$.
\end{Lem}
\begin{proof}
To see that $\alpha_\calA$~recognises $L(\calA)$, let
\begin{align*}
  P &:= \bigset{ I \subseteq \bbU(\TA(S_\calA)) }
               { \eta(q_0) \in I \text{ and }
                 I \text{ is upwards closed w.r.t. } {\subseteq} } \\
    &\subseteq \Branch(S_\calA)\,.
\end{align*}
For a tree $t \in \bbT_0\Sigma$, it follows that
\begin{align*}
  t \in L(\calA)
  &\quad\iff\quad \text{there is a run } \varrho \text{ of } \calA \text{ on } t
                  \text{ such that } \varrho(\emptyseq) = q_0 \\
  &\quad\iff\quad \text{there is a run } \varrho \text{ of } \calA \text{ on } t
                  \text{ such that } \widetilde\pf(\varrho) = q_0 \\
  &\quad\iff\quad \eta(q_0) \in \alpha_\calA(t) \\
  &\quad\iff\quad \alpha_\calA(t) \in P\,.
\end{align*}

It remains to check that $\alpha_\calA$~is a morphism.
For a tree $t \in \bbT_n\bbT\Sigma$, we have
\begin{align*}
  \pi(\bbT\alpha_\calA(t))
  = \biglset a \bigmset {}
      & a \leq \pi(s),\ s \in^\bbT \bbT\alpha_\calA(t)\,,\ 
        \pi(s) \text{ defined} \bigrset \\
  = \biglset a \bigmset {}
      & a \leq \pi(s),\ s \simeq_\sh t,\ \pi(s) \text{ defined,} \\
      & s(v) \in \alpha_\calA(t(v)), \text{ for all } v \bigrset \\
  = \biglset a \bigmset {}
      & a \leq \pi(s),\ s \simeq_\sh t,\ \pi(s) \text{ defined, for each } v
        \text{ there is} \\
      & \text{a run } \varrho_v \text{ on } t(v) \text{ such that }
        s(v) = \eta(\widetilde\pf(\varrho_v)) \bigrset\,.
\end{align*}
We have to show that this set is equal to
\begin{align*}
  \alpha_\calA(\Flat(t))
  = \bigset{ a }{ a \leq \eta(\widetilde\pf(\varrho)) \text{ for some run } \varrho
                  \text{ on } \Flat(t) }\,.
\end{align*}

$(\supseteq)$ Let $\varrho$~be a run on $\Flat(t)$.
For $v \in \dom(t)$, let $\varrho_v$~be the restriction of~$\varrho$
to the vertices in $\dom(t(v))$ and set
$s(v) := \eta(\widetilde\pf(\varrho_v))$.
Then $\eta(\widetilde\pf(\varrho)) = \pi(s)$.

$(\subseteq)$ Let $s$~be a tree with $s(v) = \eta(\widetilde\pf(\varrho_v))$,
for some run~$\varrho_v$ on~$t(v)$.
Let $\varrho$~be the run on $\Flat(t)$ such that, for every $v \in \dom(t)$,
the restriction of~$\varrho$ to the vertices
in $\dom(t(v))$ coincides with~$\varrho_v$.
Then $\eta(\widetilde\pf(\varrho)) = \pi(s)$.
\end{proof}

\begin{Lem}\label{Lem: automaton semigroup meet-continuous}
The automaton $\omega$-semigroup~$\frakS_\calA$ is meet-continuous.
\end{Lem}
\begin{proof}
Let $U \in \bbW\calP(S)$. We have to show that
\begin{align*}
  \pi(\bbW\inf(U)) = \inf {\set{ \pi(u) }{ u \in^\bbW U }}\,.
\end{align*}
We distinguish several cases.

(1) First, suppose that $\bbW\inf(U)$ is undefined.
Then $\inf U(i)$ is undefined, for some index~$i$.
This means that $U(i)$ contains two incomparable elements.
Fix sequences $u,u' \in^\bbW U$ such that $u(i)$~and~$u'(i)$
are incomparable and $u(j) = u'(j)$, for all $j \neq i$.
If at least one of $\pi(u)$ and $\pi(u')$ is not defined,
we are done. Hence, suppose that both products are defined.
We claim that their values are incomparable and, thus, the infimum
on the right-hand side of the above equation is not defined.
For the proof, we distinguish several cases.

(1\,a) Suppose that $U(i)$ has arity~$0$.
Then $i$~is the last position.
Let $u(i) = p$ and $u'(i) = p'$.
If $i = 0$, then $\pi(u) = p \neq p' = \pi(u')$ are incomparable.
Hence, suppose that $i > 0$ and let $u(i-1) = \langle q,k,r\rangle$.
By assumption, $\pi(u)$ and $\pi(u')$ are both defined.
This implies that $p = r = p'$. A~contradiction.

(1\,b) Suppose that $U(i)$ has arity~$1$.
Let $u(i) = \langle p,k,q\rangle$ and $u'(i) = \langle p',k',q'\rangle$.
Since these values are incomparable, we have $p \neq p'$ or $q \neq q'$.

First, suppose that $p \neq p'$.
If $i > 0$ we can use the value of $u(i-1)$ to show that $p = p'$ as in Case~(1\,a) above.
A~contradiction.
Consequently, $i = 0$ and, depending on the arity,
we have either $\pi(u) = p \neq p' = \pi(u')$ or
\begin{align*}
  \pi(u) = \langle p,k,r\rangle \neq \langle p',k',r'\rangle = \pi(u')\,,
\end{align*}
for suitable $r,r' \in Q$ and $k,k' < \omega$.

Similarly, suppose that $q \neq q'$.
Again, if $i$~is not the last position,
we get a contradiction by considering the value $u(i+1)$.
It follows that
\begin{align*}
  \pi(u) = \langle r,k,q\rangle \neq \langle r',k',q'\rangle = \pi(u')\,,
\end{align*}
for suitable $r,r' \in Q$ and $k,k' < \omega$.

\smallskip
(2) It remains to consider the case where $\bbW\inf(U)$ is defined.
For every position~$i$ in the sequence~$U$, it follows that either
\begin{alignat*}{-1}
  U(i) &\subseteq \{p_i\}\,, &&\qquad\text{for some state } p_i\,, \\
\prefixtext{or}
  U(i) &\subseteq \{p_i\} \times K_i \times \{q_i\}\,, &&\qquad\text{for } p_i,q_i \in Q
  \text{ and } K_i \subseteq \omega\,.
\end{alignat*}
Hence,
\begin{align*}
  \inf U(i) = \langle p_i,k_i,q_i\rangle\,,
\end{align*}
where $k_i := \inf_\sqsubseteq K_i$ is the $\sqsubseteq$-least element of~$K_i$.
We again distinguish several cases.

(2\,a) Suppose that $q_i \neq p_{i+1}$, for some~$i$.
Then $\pi(u)$ is undefined, for all $u \in^\bbW U$, and so is $\pi(\bbW\inf(U))$.
Hence, both sides of the equation are undefined.

(2\,b) Suppose that $q_i = p_{i+1}$, for all~$i$, and the sequence~$U$ is infinite.
For every $u \in^\bbW U$, we have
$u(i) = \langle p_i,m_i,q_i\rangle$, for some $m_i \in K_i$.
Consequently, $\liminf_i k_i \sqsubseteq \liminf_i m_i$.

If $\liminf_i k_i$~is even, so is $\liminf_i m_i$.
This implies that all products $\pi(u)$ are defined and so is $\pi(\bbW\inf(U))$.
Consequently, $\pi(u) = p_0 = \pi(\bbW\inf(U))$.

If $\liminf_i k_i$~is odd, $\pi(\bbW\inf(U))$~is undefined.
Choosing $u \in^\bbW U$ with $u(i) = \langle p_i,k_i,q_i\rangle$,
it follows that $\pi(u)$ and, therefore, the infimum on the right-hand side of the equation
is also undefined.

(2\,c) Suppose that $q_i = p_{i+1}$, for all~$i$, and the sequence~$U$ has length $n < \omega$.
If the last element of~$U$ has arity~$0$, then
$\pi(\bbW\inf(U)) = p_0$ and $\pi(u) = p_0$, for all $u \in^\bbW U$.
Otherwise, we have $\pi(\bbW\inf(U)) = \langle p_0,\inf_i k_i,q_{n-1}\rangle$ and,
for $u \in^\bbW U$ with $u(i) = \langle p_i,m_i,q_i\rangle$,
$\pi(u) = \langle p_0,\inf_i m_i,q_{n-1}\rangle$ where $k_i \sqsubseteq m_i$.
As above, we can chose $u \in^\bbW U$ with $m_i = k_i$.
Consequently, the infimum on the right-hand side also evaluates to
$\langle p_0,\inf_i k_i,q_{n-1}\rangle$.
\end{proof}

\begin{Thm}\label{Thm: branch-continuity captures regularity}
Let $\Sigma$~be a finite alphabet and $L \subseteq \bbT_0\Sigma$.
The following statements are equivalent.
\begin{enum1}
\item $L$~is $\MSO$-definable.
\item $L$~is recognised by some nondeterministic parity automaton.
\item $L$~is recognised by a morphism $\varphi : \bbT\Sigma \to \Branch(\frakS)$
  for some finite, meet-continuous $\omega$-semigroup\/~$\frakS$.
\item $L$~is recognised by some morphism $\varphi : \bbT\Sigma \to \frakA$
  to a finitary, branch-con\-tinu\-ous tree algebra~$\frakA$.
\end{enum1}
\end{Thm}
\begin{proof}
(1)~$\Leftrightarrow$~(2) is standard\?;
the implication (4)~$\Rightarrow$~(1) was proved in
Proposition~\ref{Prop: recognisable languages are definable}\?;
and
(3)~$\Rightarrow$~(4) holds by Theorem~\ref{Thm: Branch(S) branch-continuous}.
Finally, the implication (2)~$\Rightarrow$~(3) follows by
Lemmas \ref{Lem: alpha recognises L}~and~\ref{Lem: automaton semigroup meet-continuous}.
\end{proof}

\section{RT-algebras}   

\subsection{Regular trees and unravellings}   

When we want to compute tree algebras we have to represent them in a finite way.
Even for a finitary algebra, two problems arise\?:
there are infinitely many sorts and the product $\pi : \bbT A \to A$ has an infinite
domain.
In this section, we look at finite representations of the product function.
We start by looking at algebras where the product is only defined for \emph{regular}
trees. Such algebras correspond to Wilke algebras in the semigroup setting.
For lack of a better name, we will call them \emph{RT-algebras.}
(The term `regular tree algebra' is unfortunately already taken.)
\begin{Def}
(a) We denote by $\bbT^\reg_n A$ the subset of $\bbT_n A$ consisting of all
\emph{regular} trees.

(b)
$\frakA = \langle A,\pi,{\leq}\rangle$ is an \emph{RT-algebra}
if $\pi : \bbT^\reg A \to A$ is a $\bbT^\reg$-algebra.

(c) The \emph{regular restriction} of a tree algebra
$\frakA = \langle A,\pi,{\leq}\rangle$ is the corresponding RT-algebra
\begin{align*}
  \frakA^\reg := \langle A,\ \pi \restriction \bbT^\reg A,\ {\leq}\rangle\,.
\end{align*}
\upqed
\markenddef
\end{Def}

Note that RT-algebras can be seen as a particular form of partial tree algebras.
Hence, many definitions and theorems about tree algebras apply.
Furthermore,
properties of tree algebras that are defined solely in terms of the order and
finite products directly transfer from a tree algebra to the corresponding RT-algebra.
For examples, the algebras $\frakA$~and~$\frakA^\reg$ have the same
cylinder maps and the same sets of join-generators.

One way to define regular trees is as unravellings of finite graphs.
As we are dealing with trees where the successors are ordered
from left-to-right, we need to do the same in our graphs.
For this reason we label the edges by natural numbers to distinguish
the successors of a vertex.
\begin{Def}
Let $A$~be a ranked set.

(a) An \emph{$A$-labelled graph} $\frakG = \langle V,E,\lambda,\eta,v_0\rangle$
consists of a directed graph $\langle V,E\rangle$ with a distinguished
\emph{root vertex} $v_0 \in V$ and two labelling functions
$\lambda : V \to A$ and $\eta : E \to \omega$ such that
every vertex $v \in V$ has exactly $n := \ar(\lambda(v))$ outgoing edges
$e_0,\dots,e_{n-1}$ and their labels are $\eta(e_i) = i$, for $i < n$.
We call the end vertex of~$e_i$ the \emph{$i$-th successor} of~$v$.

(b) The \emph{unravelling} $\un(\frakG)$ of an $A$-labelled graph
$\frakG = \langle V,E,\lambda,\eta,v_0\rangle$
is the $A$-labelled tree whose vertices are all paths of~$\frakG$ that start at the
root~$v_0$ and each such path is labelled by the label in~$\frakG$ of its end vertex.
For two graphs $\frakG$~and~$\frakH$, we write $\frakG \simeq_\un \frakH$
if they have the same unravelling.

(c) We denote by $\bbG_n A$~the set of all finite graphs
whose unravelling is a tree in $\bbT_n A$.
Let $\un_A : \bbG A \to \bbT A$
be the function mapping each graph to its unravelling and let
$\Flat_A : \bbG\bbG A \to \bbG A$
be the flattening function for graphs (which is defined in the natural way).

(d) For $G,G' \in \bbG A$, we write $G \simeq_\sh G'$ if these graphs
only differ in the vertex labelling with respect to~$A$, i.e.,
they have the same sets of vertices and edges
and the same vertices are labelled by variables~$x_i$.
\markenddef
\end{Def}

\begin{Rem}
$\un(\Flat(G)) = \Flat(\un(\bbG\un(G)))\,,
 \quad\text{for all } G \in \bbG\bbG A\,.$
\begin{center}
\includegraphics{Algebras-submitted-13.mps}
%
%
%
%
%
\end{center}
\end{Rem}

For most regular trees are the unravelling of several graphs.
The following technical results help us in choosing a convenient one.
\begin{Lem}\label{Lem: combining regular trees}
For all $G \in \bbG_0 A$ and $H \in \bbG_0 B$, there exist graphs
$G' \in \bbG_0 A$ and $H' \in \bbG_0 B$ such that
\begin{align*}
  G \simeq_\un G' \simeq_\sh H' \simeq_\un H\,.
\end{align*}
\end{Lem}
\begin{proof}
The direct product $K := G \times H$ is a finite $(A \times B)$-labelled graph.
Let $G'$~and~$H'$ be the graphs obtained from~$K$ by projecting the labels
to their two components.
Then $G \simeq_\un G'$ and $H \simeq_\un H'$.
Furthermore, $G' \simeq_\sh H'$.
\end{proof}

\begin{Cor}
Let $t_0,\dots,t_m \in \bbT^\reg_n A$ with $m,n < \omega$.
If $t_0 \simeq_\sh\dots\simeq_\sh t_m$,
there exist finite graphs $G_0 \simeq_\sh\dots\simeq_\sh G_m$ such that
$t_i = \un(G_i)$, for all $i \leq m$.
\end{Cor}
\begin{proof}
As each tree~$t_i$ contains only finitely many variables,
we can decompose it as $t_i = p_i(s^i_0,\dots,s^i_l)$ where
$p_i$~is a finite tree and each $s^i_k$~either is a tree without variables or
$s^i_k = \sing(x_j)$, for some variable~$x_j$.
Since $t_0 \simeq_\sh\dots\simeq_\sh t_m$, we can choose these trees
such that
\begin{align*}
  p_0 \simeq_\sh\dots\simeq_\sh p_m
  \qtextq{and}
  s^0_k \simeq_\sh\dots\simeq_\sh s^m_k\,,  \quad\text{for all } k \leq l\,.
\end{align*}
For those~$k$ where $s^i_k$~does not contain variables, we can use
Lemma~\ref{Lem: combining regular trees} to find finite graphs
\begin{align*}
  H^0_k \simeq_\sh\dots\simeq_\sh H^m_k
  \qtextq{with}
  \un(H^i_k) = s^i_k\,.
\end{align*}
For indices~$k$ with $s^i_k =\nobreak \sing(x_j)$, we choose for~$H^i_k$
the singleton graph whose only vertex is labelled~$x_j$.
Then the graphs $G_i := p_i(H^i_0,\dots,H^i_l)$ have the desired property.
\end{proof}

\subsection{Traces and regularisations}   

Our next goal is to prove that finitary, branch-continuous tree algebras are
determined by their regular restrictions.
For the proof, we will use the representation of a branch-continuous tree algebra
as a quotient of an algebra of the form $\Branch(\frakS) = \bbD\bbU(\TA(\frakS))$.
We start by recovering the $\omega$-semigroup~$\frakS$ from~$\frakA^\reg$.
In semigroup theory there is a standard way to expand a so-called
\emph{Wilke algebra,} the analogue of an RT-algebra,
to an $\omega$-semigroup.
\begin{Def}
(a) The functor $\bbW^\reg : \pPos \to \pPos$ is defined by
\begin{align*}
  \bbW^\reg_0 A &:= A_1^{<\omega}A_0 \cup
                    \set{ w \in A_1^\omega }{ w \text{ ultimately periodic} }\,, \\
  \bbW^\reg_1 A &:= A_1^{<\omega}, \\
  \bbW^\reg_n A &:= \emptyset\,, \quad\text{for } n > 1\,.
\end{align*}

(b) An ordered \emph{Wilke algebra} $\langle A,\pi,{\leq}\rangle$ is a
$\bbW^\reg$-algebra $\pi : \bbW^\reg A \to A$.

(c) Given an $\omega$-semigroup~$\frakS$, we denote by $\frakS^\reg$ the corresponding
Wilke algebra.
\markenddef
\end{Def}
The following is a standard result in the theory of $\omega$-semigroups
(see, e.g., Theorem~II.5.1 of~\cite{PerrinPin04}).
\begin{Thm}\label{Thm: Wilke algebras}
\itm{(a)}
For every finite Wilke algebra\/~$\frakS_0$, there exists a unique
$\omega$-semi\-group\/~$\frakS$ such that\/ $\frakS^\reg = \frakS_0$.

\itm{(b)}
Every morphism $\varphi : \frakS_0 \to \frakT_0$ between finite Wilke
algebras is also a morphism $\varphi : \frakS \to \frakT$ between
the corresponding $\omega$-semigroups.
\end{Thm}

We will use this theorem to recover the trace $\omega$-semi\-group from a
RT-algebra.
\begin{Def}
Let $\frakA$~be an RT-algebra.

(a) The \emph{Wilke algebra $\SG^\reg(\frakA)$ associated} with~$\frakA$ is the
Wilke algebra with domains $A_0$~and~$A_1$ whose product is inherited from that
of~$\frakA$.

(b) If $\frakA$~is finitary, we define the \emph{$\omega$-semigroup $\SG(\frakA)$
associated} with~$\frakA$ as the unique $\omega$-semigroup whose associated
Wilke algebra is equal to $\SG^\reg(\frakA)$.
\markenddef
\end{Def}

\begin{Prop}\label{Prop: TS the same in A and A^reg}
Let\/ $\frakA$~be a finitary tree algebra and $S \subseteq A_0 \cup A_1$.
\begin{enuma}
\item $\SG(\frakA) = \SG(\frakA^\reg)\,.$
\item When computing $\langle S\rangle$ in\/ $\frakA$ and\/~$\frakA^\reg$,
  we obtain the same result.
\item $\frakB$~is a semigroup-like subalgebra of\/~$\frakA$ if,
  and only if, $\frakB^\reg$~is a semi\-group-like subalgebra of\/~$\frakA^\reg$.
\item For every finitary, semigroup-like RT-algebra\/~$\frakB_0$,
  there exists a unique semi\-group-like tree algebra\/~$\frakB$\/ with
  $\frakB^\reg = \frakB_0$.
\end{enuma}
\end{Prop}
\begin{proof}
(a) follows from the fact that both $\omega$-semigroups have the same associated
Wilke algebra.

(d) follows by Theorem~\ref{Thm: Wilke algebras} and the fact
that every semigroup-like tree algebra~$\frakB$ is uniquely determined by
its associated $\omega$-semigroup $\SG(\frakB)$.

(c) follows by~(b).

(b) Let $C$~be the result when computing $\langle S\rangle$ in~$\frakA^\reg$
and let $D$~be the result when computing it in~$\frakA$.
Then
\begin{align*}
  C = \rng \pi \restriction \bbT^\reg S
  \qtextq{and}
  D = \rng \pi \restriction \bbT S\,.
\end{align*}
Since $\bbT^\reg S \subseteq \bbT S$, it follows that $C \subseteq D$.

For the converse, note that $\pi$~induces an associative function
\begin{align*}
  \tilde\pi : \bbW(D_0 \cup D_1) \to D_0 \cup D_1\,,
\end{align*}
i.e., an $\omega$-semigroup $\frakD = \langle \tilde D,\tilde\pi\rangle$
with $\tilde D := D_0 \cup D_1$.
In the same way, we obtain a Wilke algebra
$\frakC^\reg = \langle C',\tilde\pi_0\rangle$ where $C' := C_0 \cup C_1$
and $\tilde\pi_0 : \bbW^\reg C' \to C'$.
Let $\frakC = \langle C',\tilde\pi_1\rangle$ be the $\omega$-semigroup associated
with~$\frakC^\reg$.
As every element of~$C'$ can be written as a regular product of elements of~$S$,
it follows that $\frakC^\reg$~and, thus, $\frakC$~are generated by~$S$.
In the same way, we see that $\frakD$~is generated by~$S$.
Consequently, $D' = C'$, which implies that $D = C$.
\end{proof}

\begin{Def}
Let $\frakA$~be a finitary RT-algebra, $\frakS \subseteq \frakA$ a
semigroup-like subalgebra, and $t \in \bbT A$ a tree.

(a) The \emph{trace set} of~$t$ is
\begin{align*}
  \Tr_S(t) := \Aboveseg \set{ \pi(u) }{ u \text{ an $S$-trace of } t }\,.
\end{align*}

(b) An \emph{$S$-regularisation} of~$t$ is a regular tree $t_0 \in \bbT^\reg A$ such that
\begin{align*}
  \Tr_S(t_0) = \Tr_S(t)\,,
\end{align*}
and every label used by~$t_0$ also occurs somewhere in~$t$.
\markenddef
\end{Def}

\begin{Rem}
If the tree algebra~$\frakA$ is branch-continuous with skeleton~$\frakS$,
it follows by Lemma~\ref{Lem: product in branch-continuous algebras} that
\begin{align*}
  \pi(t) = \inf \Tr_S(t)\,, \quad\text{for all } t \in \bbT(\cl(S))\,.
\end{align*}
\end{Rem}

As a first application of trace sets, we prove that
every finitary RT-algebra can be expanded to
a branch-continuous tree algebra in at most one way.
\begin{Lem}\label{Lem: at most extension to a T-algebra}
Let\/ $\frakA$~and\/~$\frakB$ be two finitary, branch-continuous tree algebras
with skeletons $\frakS \subseteq \frakA$ and $\frakT \subseteq \frakB$. Then
\begin{align*}
  \frakA^\reg = \frakB^\reg
  \qtextq{and}
  S = T
  \qtextq{implies}
  \frakA = \frakB\,.
\end{align*}
\end{Lem}
\begin{proof}
Let $\pi : \bbT A \to A$ be the product of~$\frakA$ and
$\pi' : \bbT A \to A$ the product of~$\frakB$.
By Proposition~\ref{Prop: TS the same in A and A^reg}\,(d),
$\pi$~and~$\pi'$ agree on trees in~$\bbT S$.
As the orderings of $\frakA$~and of~$\frakB$ also coincide,
it follows that the closure $C := \cl(S)$ is the same in both algebras.
Finally, note that the definition of $\Tr_S(t)$ only depends on
$\frakA^\reg = \frakB^\reg$.
For a tree $t \in \bbT C$, it therefore follows by
Lemma~\ref{Lem: product in branch-continuous algebras} that
\begin{align*}
  \pi(t) = \inf \Tr_S(t) = \inf \Tr_T(t) = \pi'(t)\,.
\end{align*}
For an arbitrary term $t \in \bbT A$, we then have
\begin{align*}
  \pi(t) &= \sup {\set{ \pi(s) }{ s \in \bbT C\,,\ s \leq^\bbT t }} \\
         &= \sup {\set{ \pi'(s) }{ s \in \bbT C\,,\ s \leq^\bbT t }}
          = \pi'(t)\,.
\end{align*}
\upqed
\end{proof}

To prove the existence of regularisations, we employ a result
from~\cite{Blumensath13a} on additive labellings.
\begin{Def}
(a) Let $\frakG = \langle V,E,v_0\rangle$ be a graph with a distinguished
root vertex~$v_0$, let $L \subseteq V$ be the set of leaves of~$\frakG$,
and let $\frakS$~be an $\omega$-semigroup.
An \emph{additive labelling} of~$\frakG$ is a function
$\lambda : E \cup L \to S$ mapping edges of~$\frakG$ to unary elements
and leaves to $0$-ary elements.

(b)
For an additive labelling~$\lambda$ of~$\frakG$ and a (finite or infinite)
path $\beta = (e_n)_n$ of~$\frakG$, we define
\begin{align*}
  \lambda(\beta) := \prod_n \lambda(e_n)\,.
\end{align*}
(If $\beta = e_0\dots e_m$ is finite in the above definition,
we allow the last element~$e_m$ to be a leaf instead of an edge.)
If $\frakG$~is a tree and $x \prec y$ are vertices of~$\frakG$,
we also write
\begin{align*}
  \lambda(x,y) := \lambda(\beta)\,,
  \quad\text{where $\beta$~is the unique path from~$x$ to~$y$.}
\end{align*}

(b) The \emph{limit set} of~$\lambda$ is
\begin{align*}
  \lim \lambda :=
    \set{ \lambda(\beta) }{ \beta \text{ a maximal path of } \frakG
                            \text{ starting at the root} }\,.
\end{align*}
\upqed
\markenddef
\end{Def}

The following has been proven in~\cite{Blumensath13a}.
\begin{Thm}\label{Thm: trees to regular trees}
Let $\lambda$~be an additive labelling of a tree~$t$.
There exists a finite graph~$G$ and an additive labelling~$\lambda'$ of~$G$
such that
\begin{align*}
  \lim \lambda = \lim \lambda'
  \qtextq{and}
  \rng \lambda' \subseteq \rng \lambda\,.
\end{align*}
\end{Thm}
We also need a version for regular trees.
\begin{Thm}\label{Thm: regular trees to regular trees}
Let $\lambda$~be an additive labelling of a regular tree~$t$.
There exists a finite graph~$G$ and an additive labelling~$\lambda'$ of~$G$
such that
\begin{align*}
  \lim \lambda = \lim \lambda'\,,\quad
  \rng \lambda' \subseteq \rng \lambda\,,
  \qtextq{and}
  \un(G) \simeq_\sh t\,.
\end{align*}
\end{Thm}
\begin{proof}
(This proof uses terminology and notation from~\cite{Blumensath13a}.)
Let $H$~be a finite graph such that $t = \un(H)$ and let $p : \dom(t) \to \dom(H)$
be the corresponding graph homomorphism.
We fix a bijection $\eta : \dom(H) \to [n]$, for some $n < \omega$.
Given a Ramseyan split~$\sigma$ of~$\lambda$, we define a function~$\sigma'$ by
\begin{align*}
  \sigma'(v) := n \cdot \sigma(v) + \eta(p(v))\,.
\end{align*}
Since $u \sqsubset_{\sigma'} v$ implies $u \sqsubset_\sigma v$, it follows that
$\sigma'$~is also a Ramseyan split of~$\lambda$.
Let $P \subseteq \dom(t)$ be a set such that
\begin{align*}
  \lim \lambda^P_{\sigma'} = \lim \lambda\,.
\end{align*}

We claim that the graph $G := \frakC^P_{\sigma'}(\lambda)$ and the labelling
$\lambda' := \lambda^P_{\sigma'}$ have the desired properties.
The inclusion $\rng \lambda' \subseteq \rng \lambda$ holds by definition
of~$\lambda'$, and the equation $\lim \lambda' = \lim \lambda$ by choice of~$P$.
For the second statement note that, by definition of~$\sigma'$,
there exists a graph homomorphism $\varphi : \dom(G) \to \dom(H)$
(ignoring the labelling) which extends to the corresponding unravellings.
Consequently, $\un(G) \simeq_\sh \un(H) = t$.
\end{proof}
\begin{Rem}
In both of the above theorems we can also bound the length of the longest path contained
in the graph~$G$. This bound only depends on the size of the $\omega$-semigroup used
by~$\lambda$ and, in the second statement, also on the size of the graph~$H$.
It does not depend on~$t$.
\end{Rem}

We use these two theorems to prove the existence of regularisations.
To do so, we have to construct suitable additive labellings.
\begin{Lem}\label{Lem: additive labelling for traces}
Let\/ $\frakA$~be a finitary RT-algebra, $\frakS \subseteq \frakA$ a
semigroup-like subalgebra, and $n < \omega$.
There exists a finite $\omega$-semigroup\/~$\frakT$ and a partial function
$f : T \to \PSet(A_n)$ such that every tree $t \in \bbT_n(\cl(S))$ has an additive
labelling~$\lambda_t$ over\/~$\frakT$ with
\begin{align*}
  \Tr_S(t) = \bigcup f[\lim \lambda_t]\,.
\end{align*}
\end{Lem}
\begin{proof}
Set $\frakT' := \SG(\frakS)$.
The desired $\omega$-semi\-group~$\frakT$ is derived from the
tree algebra $\bbU\TA(\frakT')$ as follows. The domains are
\begin{align*}
  T_0 := \bbU T'_0 \cup \bigl((\bbU T'_1 \cup \{1\}) \times [n]\bigr)
  \qtextq{and}
  T_1 := \bbU T'_1\,.
\end{align*}
The product of~$\frakT$ extends that of~$\bbU\frakT'$ by
\begin{align*}
  I\cdot\langle J,k\rangle := \langle IJ,k\rangle\,,
  \quad\text{for } I \in \bbU T'_1 \text{ and }
  \langle J,k\rangle \in (\bbU T'_1 \cup \{1\}) \times [n]
\end{align*}
(where $I\cdot 1 := I$).
We use the partial function $f : T \to \PSet(A_n)$ defined by
\begin{alignat*}{-1}
  f(I) &:= I\,, &&\qquad\text{for } I \in \bbU T'_0 \cup \bbU T'_1\,, \\
  f(\langle J,k\rangle) &:= J(x_k)\,,
    &&\qquad\text{for } \langle J,k\rangle \in \bbU T'_1 \times [n]\,, \\
  f(\langle 1,k\rangle) &\text{ is undefined.}
\end{alignat*}

Finally, given a tree $t \in \bbT_n(\cl(S))$,
we define the desired additive labelling~$\lambda_t$ over~$\frakT$ by
\begin{alignat*}{-1}
  \lambda_t(v,vk) &:= \set{ c \in S_0 \cup S_1 }{ \cy_k(c) \geq t(v) }\,, \mkern-54mu\\
  & &&\text{for } v \in \dom(t) \text{ and } k < \ar(t(v)), \\
  \lambda_t(v) &:= \set{ c \in S_0 }{ c \geq t(v) }\,,
    \quad&&\text{for leaves } v \in \dom(t) \setminus \Hole(t), \\
  \lambda_t(v) &:= \langle 1,k\rangle\,,
    \quad&&\text{for holes } v = \hole_k(t).
\end{alignat*}
Then it follows for an $S$-trace~$u$ of~$t$ along some branch~$\beta$ that
\begin{align*}
  u(0^n) \in \lambda_t(\beta \restriction n,\ \beta \restriction (n+1))\,.
\end{align*}
Hence,
\begin{align*}
  \pi(u) \in \bigcup f(\lambda_t(\beta))\,.
\end{align*}
Consequently, $\Tr_S(t) = \bigcup f[\lim \lambda_t]$.
\end{proof}

\begin{Thm}\label{Thm: existence of regularisations}
Let\/ $\frakA$~be a finitary RT-algebra and $\frakS \subseteq \frakA$ a
semigroup-like subalgebra.
\begin{enuma}
\item Every tree $t \in \bbT A$ has an $S$-regularisation $t_0 \in \bbT^\reg A$.
\item If there exists some function $f : A \to B$ such that
  $\bbT f(t)$~is regular, then we can choose the $S$-regularisation~$t_0$
  such that $t_0 \simeq_\sh t$.
\end{enuma}
\end{Thm}
\begin{proof}
Given a tree $t \in \bbT_n A$, we use the labelling~$\lambda_t$ over the
$\omega$-semigroup~$\frakT$ from Lemma~\ref{Lem: additive labelling for traces}
to find an $S$-regularisation of~$t$ as follows.
By Theorem~\ref{Thm: trees to regular trees},
there exists a finite graph~$G$ and an additive labelling~$\lambda_G$ of~$G$
such that $\lim \lambda_G = \lim \lambda_t$ and
$\rng \lambda_G \subseteq \rng \lambda_t$.
For~(b), we can use Theorem~\ref{Thm: regular trees to regular trees}
to ensure that $\un(G) \simeq_\sh t$.
Let $t_0 \in \bbT^\reg_n A$ be a regular tree such that
$t_0 \simeq_\sh \un(G)$ and the labelling~$\lambda_{t_0}$ associated with~$t_0$
coincides with the (unravelling of)~$\lambda_G$.
Then
\begin{align*}
  \Tr_S(t) = \bigcup f[\lim \lambda_t] = \bigcup f[\lim \lambda_{t_0}] = \Tr_S(t_0)\,.
\end{align*}
Moreover, in case~(b) we have $t_0 \simeq_\sh \un(G) \simeq_\sh\nobreak t$.

Hence, it remains to prove that every label used in~$t_0$ also occurs in~$t$.
The construction above does not yield this fact. We have to modify it slightly
by changing the labelling~$\lambda$ such that the value $\lambda(x,y)$
also encodes the label $t(x)$ (say, by using a suitable $\omega$-semigroup with
domain $A \times T$ and setting $\lambda'_t(x,y) := \langle t(x),\lambda_t(x,y)\rangle$).
Then the claim follows from the condition that $\rng \lambda_G \subseteq \rng \lambda_t$.
\end{proof}

Existence of regularisations can be strengthened in the following way.
\begin{Lem}\label{Lem: regularisations satisfying a relation}
Let\/ $\frakA$~and\/~$\frakB$ be two finitary, branch-continuous tree algebras,\/
$\frakS \subseteq \frakA$ and\/ $\frakT \subseteq \frakB$ corresponding skeletons,
and $\theta \subseteq A \times B$ a binary relation.
For every pair of trees $s \in \bbT A$ and $t \in \bbT B$ with
$s \mathrel{\theta^\bbT} t$,
there are an $S$-regularisation~$s_0$ of~$s$
and a $T$-regularisation~$t_0$ of~$t$
such that $s_0 \mathrel{\theta^\bbT} t_0$.
\end{Lem}
\begin{proof}
$s \mathrel{\theta^\bbT} t$ implies $s \simeq_\sh t$.
Hence, there exists a tree $u \in \bbT(A \times B)$ such that
$s = \bbT p(u)$ and $t = \bbT q(u)$, where $p : A \times B \to A$ and
$q : A \times B \to B$ are the two projection functions.
By Theorem~\ref{Thm: branch-continuity preserved by products}, the product
$\frakA \times \frakB$ is finitary and branch-continuous with skeleton
$\frakS \times \frakT$.
Consequently, we can use Theorem~\ref{Thm: existence of regularisations}
to find an $(S \times T)$-regularisation~$u_0$ of~$u$.
Set $s_0 := \bbT p(u_0)$ and $t_0 := \bbT q(u_0)$.
Then
\begin{alignat*}{-1}
  \Tr_S(s_0) &= p[\Tr_S(u_0)] &&= p[\Tr_S(u)] &&= \Tr_S(s) \\
\prefixtext{and}
  \Tr_S(t_0) &= q[\Tr_S(u_0)] &&= q[\Tr_S(u)] &&= \Tr_S(t)\,.
\end{alignat*}
Hence, $s_0$~is an $S$-regularisation of~$s$ and $t_0$~is a $T$-regularisation of~$t$.
Furthermore,
\begin{align*}
  s(v) \mathrel\theta t(v)
  \qtextq{implies}
  u(v) \in \theta\,, \quad\text{for all } v \in \dom(u)\,.
\end{align*}
As all labels used by~$u_0$ also appear in~$u$, we have
\begin{align*}
  u_0(v) \in \theta\,,
  \qtextq{which implies that}
  s_0(v) \mathrel\theta t_0(v)\,, \quad\text{for all } v \in \dom(u)\,.
\end{align*}
Consequently, $s_0 \mathrel{\theta^\bbT} t_0$.
\end{proof}

\subsection{Expansion of the regular product}   

We have already shown in Lemma~\ref{Lem: at most extension to a T-algebra} that
an RT-algebra can be expanded to at most one full tree algebra.
In general such an expansion does not need to exist, but it does in the case
of algebras that are finitary and branch-continuous.
Let us start by defining branch-continuity for RT-algebras.

\begin{Def}
A RT-algebra~$\frakA$ is \emph{branch-continuous} if it is complete,
distributive, and it has a semigroup-like subalgebra $\frakS \subseteq \frakA$
with the following properties.
\begin{itemize}
\item $C := \cl(S)$ is a set of join-generators of~$\frakA$.
\item $\SG(\frakS)$ is meet-continuous.
\item For every tree $U \in \bbT\PSet(S)$,
  \mathindent=1em%
  \begin{align*}
    \sup {\bigset{ \inf \Tr_S(s) }{ s \in \bbT C\,,\ s \leq^\bbT \bbT\inf(U) }}
  = \inf {\bigset{ \inf \Tr_S(s) }{ s \in^\bbT U }}\,.
  \end{align*}
\item For every regular tree $t \in \bbT^\reg A$,
  \mathindent=1em%
  \begin{align*}
    \pi(t) = \sup {\bigset{ \inf \Tr_S(s) }
                          { s \in \bbT C \text{ with } s \leq^\bbT t }}\,.
  \end{align*}
\item For arbitrary trees $U,U' \in \bbT\bbD C$ with
  $\bbT\sup(U) = \bbT\sup(U')$,
  \mathindent=1em%
  \begin{align*}
       \sup {\bigset{ \inf \Tr_S(s) }{ s \in^\bbT U }}
     = \sup {\bigset{ \inf \Tr_S(s') }{ s' \in^\bbT U' }}\,.
  \end{align*}
\item For arbitrary trees $U,U' \in \bbT\bbU S$ with
  $\bbT\inf(U) = \bbT\inf(U')$,
  \mathindent=1em%
  \begin{align*}
       \inf {\bigset{ \inf \Tr_S(s) }{ s \in^\bbT U }}
     = \inf {\bigset{ \inf \Tr_S(s') }{ s' \in^\bbT U' }}\,.
  \end{align*}
\end{itemize}
Such a subalgebra~$\frakS$ is called a \emph{skeleton} of~$\frakA$.
\markenddef
\end{Def}

\begin{Lem}\label{Lem: morphisms of regular algebras transfer rect. bases}
Let $\varphi : \frakA \to \frakB$ be a surjective morphism of
RT-algebras that preserves arbitrary meets and joins.
\begin{enuma}
\item If\/ $\frakS$ is a skeleton of\/~$\frakA$,
  then $\varphi[\frakS]$ is a one of\/~$\frakB$.
\item If\/ $\frakA$~is branch-continuous, then so is\/~$\frakB$.
\end{enuma}
\end{Lem}
\begin{proof}
(a)
Let $T := \varphi[S]$ and $D := \varphi[C]$ where $C := \cl(S)$.
All conditions in the definition of a skeleton take the form of an
equation between terms involving meets, joins, and products.
Every equation of this form is preserved by~$\varphi$.

(b) According to Lemma~\ref{Lem: surjective morphisms preserve completeness},
$\frakB$~is complete and distributive.
Hence, the claim follows by~(a).
\end{proof}

\begin{Lem}\label{Lem: transfer of branch-continuity}
Let\/ $\frakA$~be a finitary, branch-continuous tree algebra and\/
$\frakS$~a skeleton of\/~$\frakA$.
Then $\frakA^\reg$ is branch-continuous and $\frakS^\reg$~is a skeleton of~$\frakA^\reg$.
\end{Lem}
\begin{proof}
First note that $\frakA^\reg$ is complete and distributive
since these two properties are defined solely in terms of the ordering.
Hence, it remains to prove that $\frakS^\reg$~is a skeleton of~$\frakA^\reg$.
Clearly, the set $C := \cl(S)$ is a set of join-generators.
Hence, it remains to prove the following ones.
\begin{enuma}
\item $\TS(S)$ is meet-continuous.
\item For every tree $U \in \bbT\PSet(S)$,
  \mathindent=1em%
  \begin{align*}
    \sup {\bigset{ \inf \Tr_S(s) }{ s \in \bbT C\,,\ s \leq^\bbT \bbT\inf(U) }}
  = \inf {\bigset{ \inf \Tr_S(s) }{ s \in^\bbT U }}\,.
  \end{align*}
\item For every regular tree $t \in \bbT^\reg A$,
  \mathindent=1em%
  \begin{align*}
    \pi(t) = \sup {\bigset{ \inf \Tr_S(s) }
                          { s \in \bbT C \text{ with } s \leq^\bbT t }}\,.
  \end{align*}
\item For arbitrary trees $U,U' \in \bbT\bbD C$ with
  $\bbT\sup(U) = \bbT\sup(U')$,
  \mathindent=1em%
  \begin{align*}
       \sup {\bigset{ \inf \Tr_S(s) }{ s \in^\bbT U }}
     = \sup {\bigset{ \inf \Tr_S(s') }{ s' \in^\bbT U' }}\,.
  \end{align*}
\item For arbitrary trees $U,U' \in \bbT\bbU S$ with
  $\bbT\inf(U) = \bbT\inf(U')$,
  \mathindent=1em%
  \begin{align*}
       \inf {\bigset{ \inf \Tr_S(s) }{ s \in^\bbT U }}
     = \inf {\bigset{ \inf \Tr_S(s') }{ s' \in^\bbT U' }}\,.
  \end{align*}
\end{enuma}

(a) follows from Proposition~\ref{Prop: TS the same in A and A^reg}
and the fact that $\frakS$~is meet-continuously embedded in~$\frakA$.

For (b), let $U \in \bbT\PSet(S)$.
Since $\frakS$ is meet-continuously embedded in~$\frakA$, we have
\begin{align*}
&         \sup {\bigset{ \inf \Tr_S(s) }
                       { s \in \bbT C\,,\ s \leq^\bbT \bbT\inf(U) }} \\
&\quad{}= \pi(\bbT\inf(U))
        = \inf {\set{ \pi(s) }{ s \in^\bbT U }}
        = \inf {\set{ \inf \Tr_S(s) }{ s \in^\bbT U }}\,.
\end{align*}

(c) follows from join-continuity of~$\frakA$ and the fact that
$\pi(s) = \inf \Tr_S(s)$, for trees $t \in \bbT C$.

For (d), consider two trees $U$~and~$U'$ as above.
Setting $t := \bbT\sup(U)$, it follows by join-continuity that
\begin{align*}
     \sup {\bigset{ \inf \Tr_S(s) }{ s \in^\bbT U }}
   = \pi(t)
   = \sup {\bigset{ \inf \Tr_S(s') }{ s' \in^\bbT U' }}\,.
\end{align*}

(e) follows in the same way as~(d) using the fact that
$\frakS$~is meet-con\-tinu\-ously embedded in~$\frakA$.
\end{proof}

We will expand a finitary, branch-continuous RT-algebra to a full
tree algebra in two steps. We first define the full product on the set
$\cl(S)$\?; then we extend it to the whole algebra.

\begin{Lem}\label{Lem: extending branch-cont. regular algebra on basis}
Let\/ $\frakA_0 = \langle A,\pi_0,{\leq}\rangle$ be a finitary,
branch-continuous RT-algebra,\/ $\frakS_0 = \langle S,\pi_0,{\leq}\rangle$
a skeleton of\/~$\frakA_0$, and let $C := \cl(S)$.
Define $\pi : \bbT C \to C$ by
\begin{align*}
  \pi(t) := \inf \Tr_S(t)\,, \quad\text{for } t \in \bbT C\,.
\end{align*}
Then\/ $\frakC := \langle C,\pi,{\leq}\rangle$ is a tree algebra
such that\/ $\frakC^\reg \subseteq \frakA_0$.
\end{Lem}
\begin{proof}
The function~$\pi$ extends~$\pi_0$ since, for a regular term $t \in \bbT^\reg C$, we have
\begin{align*}
  \pi_0(t)
  = \sup {\bigset{ \inf \Tr_S(s) }{ s \in \bbT C\,,\ s \leq^\bbT t }}
  = \inf \Tr_S(t)
  = \pi(t)\,.
\end{align*}

Hence, it remains to prove that it forms a tree algebra.
First, note that, according to Proposition~\ref{Prop: TS the same in A and A^reg},
there exists a unique semigroup-like tree algebra~$\frakS = \langle S,\pi_1,{\leq}\rangle$
with $\frakS^\reg = \frakS_0$.
Applying Proposition~\ref{Prop: reducing products to products over branches}
to the tree algebra~$\frakS$, it further follows that its product $\pi_1 : \bbT S \to S$
takes the form
\begin{align*}
  \pi(t) = \inf \Tr_S(t)\,, \quad\text{for } t \in \bbT S\,.
\end{align*}
One of the axioms of a skeleton states that this function~$\pi_1$ satisfies the
meet-extension condition.
Therefore, we can apply Proposition~\ref{Prop: extending product from meet-generators}
to the embedding $S \to A$,
and it follows that there exists a tree algebra
$\frakC = \langle C,\pi',{\leq}\rangle$ where the product
$\pi' : \bbT C \to C$ extends~$\pi_0$ and it is given by
\begin{align*}
  \pi'(t) := \inf {\set{ \pi_1(s) }{ s \in \bbT S\,,\ s \geq^\bbT t }}\,.
\end{align*}
We claim that $\pi' = \pi$. For $t \in \bbT C$, we have
\begin{align*}
  \pi'(t)
  &= \inf {\bigset{ \pi_1(s) }{ s \in \bbT S\,,\ s \geq^\bbT t }} \\
  &= \inf {\bigset{ \inf \Tr_S(s) }{ s \in \bbT S\,,\ s \geq^\bbT t }} \\
  &= \inf \bigcup{\bigset{ \Tr_S(s) }{ s \in \bbT S\,,\ s \geq^\bbT t }} \\
  &= \inf {\bigset{ \pi(u) }{ s \in \bbT S\,,\ s \geq^\bbT t\,,\ 
                                u \text{ an $S$-trace of } s }} \\
  &= \inf {\set{ \pi(u) }{ u \text{ an $S$-trace of } t }} \\
  &= \inf \Tr_S(t) \\
  &= \pi(t)\,.
\end{align*}
Consequently, $\pi = \pi' : \bbT C \to C$ is a tree algebra.
\end{proof}

\begin{Prop}\label{Prop: extending branch-cont. regular algebra}
Let\/ $\frakA_0 = \langle A,\pi_0,{\leq}\rangle$ be a finitary,
branch-continuous RT-algebra and\/ $\frakS_0 = \langle S,\pi_0,{\leq}\rangle$
a skeleton of\/~$\frakA_0$.
There exists a finitary, branch-continuous tree algebra\/~$\frakA$
with\/ $\frakA^\reg = \frakA_0$.
\end{Prop}
\begin{proof}
Let $C := \cl(S)$.
In Lemma~\ref{Lem: extending branch-cont. regular algebra on basis}
we seen that the function $\pi_1 : \bbT C \to C$ with
\begin{align*}
  \pi_1(t) := \inf \Tr_S(t)\,, \quad\text{for } t \in \bbT C\,,
\end{align*}
is the product of a tree algebra extending~$\pi_0$ on~$C$.
One of the axioms of a skeleton states that
$\pi_1$~satisfies the join-extension condition.
Therefore, we can apply Proposition~\ref{Prop: extending product from join-generators}
to the embedding $C \to A$, and it follows that there exists a tree algebra
$\frakA = \langle A,\pi,{\leq}\rangle$ where
the product $\pi : \bbT A \to A$ is given by
\begin{align*}
  \pi(t) := \sup {\set{ \pi_1(s) }{ s \in \bbT C\,,\ s \leq^\bbT t }}\,.
\end{align*}

To prove that $\pi$~extends~$\pi_0$, consider a regular tree $t \in \bbT^\reg A$.
Since $S$~is a skeleton of~$\frakA_0$, we have
\begin{align*}
  \pi(t) &= \sup {\bigset{ \pi_1(s) }{ s \in \bbT C\,,\ s \leq^\bbT t }} \\
         &= \sup {\bigset{ \inf \Tr_S(s) }{ s \in \bbT C\,,\ s \leq^\bbT t }}
          = \pi_0(t)\,.
\end{align*}

We have shown that $\frakA$~is a tree algebra extending~$\frakA_0$.
Furthermore, it is clearly finitary, complete, and distributive as these properties
transfer from~$\frakA_0$.
For join-continuity, suppose that $t = \bbT\sup U$ for $t \in \bbT A$
and $U \in \bbT\PSet(A)$.
Let $U',R \simeq_\sh U$ be the trees with
\begin{align*}
  U'(v) := C \cap \Belowseg U(v)
  \qtextq{and}
  R(v) := C \cap \Belowseg t(v)\,, \quad\text{for all } v\,.
\end{align*}
Then $\sup U'(v) = \sup U(v) = t(v) = \sup R(v)$.
Since $S$~is a skeleton of~$\frakA_0$, it follows that
\begin{align*}
  \pi(t)
  &= \sup {\bigset{ \pi_1(r) }{ r \in^\bbT R }} \\
  &= \sup {\bigset{ \pi_1(r) }{ r \in^\bbT U' }} \\
  &= \sup {\bigset{ \pi_1(r) }{ r \in \bbT C,\ r \leq^\bbT s
                                   \text{ for some } s \in^\bbT U }} \\
  &= \sup {\bigset{ \sup \set{ \pi_1(r) }{ r \in \bbT C,\ r \leq^\bbT s } }
                  { s \in^\bbT U }} \\
  &= \sup {\bigset{ \pi(s) }{ s \in^\bbT U }}\,.
\end{align*}

It remains to check branch-continuity.
We claim that $S$~is a skeleton of~$\frakA$.
Clearly, $C = \cl(S)$ is a set of join-generators of~$\frakA$.
By definition of~$\pi_1$, we furthermore have
\begin{align*}
  \pi(t) = \pi_1(t) = \inf \Tr_S(t)\,, \quad\text{for } t \in \bbT C\,.
\end{align*}
Hence, we only have to show that $S$~is meet-continuously embedded in~$\frakA$.
Let $t \in \bbT A$ and $U \in \bbT\PSet(S)$ be trees such that $t = \bbT\inf(U)$.
Then
\begin{align*}
  \pi(t)
  &= \sup {\set{ \inf \Tr_S(s) }{ s \in \bbT C\,,\ s \leq^\bbT t }} \\
  &= \inf {\bigset{ \inf \Tr_S(s) }{ s \in^\bbT U }}
   = \inf {\set{ \pi(s) }{ s \in^\bbT U }}\,.
\end{align*}
\upqed
\end{proof}

Summarising our results, we have obtained the following theorem.
\begin{Thm}\label{Thm: unique expansion to full algebra}
For every finitary, branch-continuous RT-algebra~$\frakA_0$,
there exists a unique branch-continuous tree algebra~$\frakA$
with $\frakA^\reg = \frakA_0$.
\end{Thm}
\begin{proof}
Uniqueness follows by Lemma~\ref{Lem: at most extension to a T-algebra}
and existence by Proposition~\ref{Prop: extending branch-cont. regular algebra}.
\end{proof}

There is a similar statement for morphisms instead of algebras.
\begin{Prop}\label{Prop: reg morphisms are full morphisms}
Let $\varphi : \frakA^\reg \to \frakB^\reg$ be a surjective morphism
between RT-algebras that preserves meets and joins.
If\/ $\frakA^\reg$~is finitary and branch-continuous, then so is\/~$\frakB^\reg$
and $\varphi$~is a morphism\/ $\frakA \to \frakB$ between
the corresponding tree algebras.
\end{Prop}
\begin{proof}
Let $\frakS$~be a skeleton of~$\frakA^\reg$ and set $C := \cl(S)$.
According to Lemma~\ref{Lem: morphisms of regular algebras transfer rect. bases},
$\frakB^\reg$~is branch-continuous and
the image $\frakT := \varphi[\frakS]$ is a skeleton of~$\frakB^\reg$.
Hence, it remains to show that $\varphi \circ \pi = \pi \circ \bbT\varphi$.
To do so it is sufficient to prove that
\begin{align*}
  \varphi(\pi(t)) = \pi(\bbT\varphi(t))\,, \quad\text{for all } t \in \bbT C\,.
\end{align*}
Since $\bbT C$~is a set of join-generators of $\bbT\frakA$,
it then follows by Lemma~\ref{Lem: sets of join-generators}\,(b)
that $\varphi \circ \pi = \pi \circ \bbT\varphi$.

To prove the claim, let $t \in \bbT C$.
By definition of~$C$, there is a tree $T \in \bbT\bbU S$ such that $t = \bbT\inf(T)$.
Setting $T' := \bbT\bbU\varphi(T)$, it follows for $v \in \dom(T)$ that
\begin{align*}
  \inf T'(v) = \inf \Aboveseg\varphi[T(v)] = \varphi(\inf T(v)) = \varphi(t(v))\,.
\end{align*}
Moreover, note that, since $\frakS$~is semigroup-like, Theorem~\ref{Thm: Wilke algebras}
implies that
\begin{align*}
  \varphi(\pi(t)) = \pi(\bbT\varphi(t))\,, \quad\text{for all } t \in \bbT S\,.
\end{align*}
Hence, meet-continuity implies that
\begin{align*}
  \pi(\bbT\varphi(t))
  &= \inf {\set{ \pi(s') }{ s' \in^\bbT T' }} \\
  &= \inf {\set{ \pi(s') }{ s' \in^\bbT \bbT\bbU\varphi(T) }} \\
  &= \inf {\set{ \pi(\bbT\varphi(s)) }{ s \in^\bbT T }} \\
  &= \inf {\set{ \varphi(\pi(s)) }{ s \in^\bbT T }}
   = \varphi(\pi(t))\,.
\end{align*}
\upqed
\end{proof}

{\small
\bibliographystyle{siam}
\bibliography{Algebras}}

\end{document}